\documentclass[final,times,sort&compress]{elsarticle}
\usepackage{graphicx,amssymb}
\usepackage{comment}
\usepackage{amsmath}
\journal{Physica A}

\begin{document}

\begin{frontmatter}



\title{Thermal phase transitions in a mixed-spin Ising model on the Lieb lattice: Exact results beyond zero magnetic field}

\author[UPJS]{Jozef Stre\v{c}ka\corref{cor}} 
\ead{jozef.strecka@upjs.sk}
\address[UPJS]{Department of Theoretical Physics and Astrophysics, Faculty of Science, \\
P. J. \v{S}af\'{a}rik University, Park Angelinum 9, 040 01 Ko\v{s}ice, Slovak Republic}
\author[UPJS]{Katar\'ina Karl'ov\'a}
\ead{katarina.karlova@upjs.sk}
\cortext[cor]{Corresponding author.}

\begin{abstract}
We investigate the ground-state and finite-temperature properties of a mixed spin-$1/2$ and spin-$1$ Ising model on a decorated square (Lieb) lattice incorporating a uniaxial single-ion anisotropy and magnetic field. By employing the generalized decoration-iteration transformation, the model is mapped exactly onto an effective spin-$1/2$ Ising model on the square lattice characterized by an effective nearest-neighbor interaction and an effective field. The studied model consequently becomes exactly solvable even for finite values of the applied magnetic field whenever the effective field vanishes. The ground-state analysis reveals three distinct phases: ferrimagnetic phase (FRI), disordered phase (DP), and ferromagnetic (FM) phase. The ground-state boundary between FRI and DP phases gives rise to a dome-shaped surface of discontinuous thermal phase transitions, which is terminated by a line of Ising-type critical points associated with continuous thermal phase transitions. Both continuous and discontinuous thermal phase transitions belong to the exactly solvable parameter regime defined by a vanishing effective field in spite of the fact that the applied magnetic field is finite. Two consecutive discontinuous thermally-induced reentrant phase transitions DP-FRI-DP are identified in a narrow parameter region. The exact analytical predictions including reentrance, field- and thermally-driven phase transitions are independently verified by classical Monte Carlo simulations.
\end{abstract}

\begin{keyword}
Ising model \sep Lieb lattice \sep phase transitions \sep critical points \sep exact solution 
\sep reentrance
\PACS 05.50.+q  \sep 05.70.Fh \sep 75.10.Hk \sep 75.30.Kz \sep 75.40.Cx
\end{keyword}

\end{frontmatter}

\section{Introduction}
Exact solutions of interacting many-body systems play a fundamental role in statistical physics, as they provide rigorous insight into collective phenomena and serve as an invaluable benchmark for approximate analytical and numerical approaches \cite{Baxter1982}. A major milestone in this field was achieved by Onsager, who obtained the exact solution of the spin-$1/2$ Ising model on the square lattice and demonstrated the existence of a finite-temperature order-disorder phase transition \cite{Onsager1944}. This breakthrough was complemented by Yang's exact derivation of the spontaneous magnetization \cite{Yang1952}, which completed the exact analytical description of the ordered phase and established the two-dimensional Ising model as one of the cornerstones of modern statistical mechanics \cite{Baxter1982}. Despite the remarkable progress achieved over the past decades, the class of exactly solvable two-dimensional lattice-statistical spin models remains rather limited \cite{Baxter1982,Diep2013}.

An important extension of the spin-$1/2$ Ising model was independently introduced by Blume and Capel when considering the spin-1 Ising model accounting additionally for the effect of uniaxial single-ion anisotropy now commonly referred to as the Blume-Capel model \cite{Blume1966,Capel1966}. The Blume-Capel model exhibits a considerably richer thermal phase transitions compared with its spin-$1/2$ counterpart as it may undergo both discontinuous and continuous thermal phase transitions separated by a tricritical point \cite{Beale1986}. Owing to this rich critical behavior, the Blume-Capel model has attracted sustained attention and has been investigated by a wide variety of analytical and numerical approaches including mean-field theory \cite{Kaufman1990,Plascak1993}, effective-field theory \cite{Siqueira1986,Kaneyoshi1992,Yuksel2012}, renormalization-group methods \cite{Burkhardt1977,Branco1997}, series expansion \cite{Saul1974}, Monte Carlo simulations \cite{Heringa1998,Silva2006,Zukovic2013}, and rigorous analyses based on Lee-Yang zeros \cite{Biskup2000}.

Despite decades of intensive research, no general exact solution of the two-dimensional Blume-Capel model is known to date \cite{Strecka2015}. Exact analytical results are available only under highly restrictive conditions. For instance, Horiguchi established an exact solution of a more general spin-$1$ Blume-Emery-Griffiths model on a honeycomb lattice, which includes the Blume-Capel model as its special case, only within a special subspace of interaction parameters \cite{Horiguchi1986}. Wu subsequently provided an alternative derivation of this result and obtained the corresponding exact critical condition \cite{Wu1986}. Consequently, the search for new exactly solvable spin-$1$ and mixed-spin Ising systems still remains a very active topic of research in statistical mechanics, see Ref. \cite{Strecka2015} and references therein. Although general exact solutions of two-dimensional spin-1 Blume-Capel models remain elusive, a number of two-dimensional mixed-spin Ising systems are exactly tractable either directly or through exact mapping transformations \cite{Fisher1959,Syozi1972,Rojas2009,Strecka2010}. Such models have attracted considerable attention owing to their rich magnetic and critical behavior, which includes ferrimagnetic ordering, compensation phenomena, reentrant and various other types of phase transitions \cite{Goncalves1985,Goncalves1992,Jascur1998,Dakhama98,Strecka2006}. In particular, mixed-spin Ising models on decorated lattices represent a prominent class of exactly solvable spin systems, since the decorating spins can often be easily eliminated through exact mapping transformations \cite{Fisher1959,Syozi1972,Rojas2009,Strecka2010}. 

In the present work, we investigate a mixed spin-$1/2$ and spin-$1$ Ising model on a decorated square (Lieb) lattice with the uniaxial single-ion anisotropy in the presence of magnetic field. This model represents a magnetic-field extension of the model exactly solved by Ja\v{s}\v{c}ur \cite{Jascur1998} and Dakhama \cite{Dakhama98} for zero field. Notably, the Lieb lattice provides a natural framework for realizing mixed-spin systems due to its bipartite character and nontrivial geometry \cite{Lieb1989}. By employing the generalized decoration-iteration transformation \cite{Fisher1959,Syozi1972,Rojas2009,Strecka2010}, the investigated model can be mapped exactly onto an effective spin-$1/2$ Ising model on the square lattice with an effective nearest-neighbor interaction and an effective field. This procedure enables an exact determination of thermal phase transitions, magnetic, and thermodynamic properties whenever the effective field vanishes, which does not necessarily mean zero applied magnetic field. Particular attention is devoted to the interplay between the magnetic field and the uniaxial single-ion anisotropy, which gives rise to an intriguing thermally-induced reentrant phase transitions. These exact analytical results are further corroborated by classical Monte Carlo simulations, which provide independent numerical confirmation for the anticipated exact results.

The remainder of this paper is organized as follows. In Sec.~\ref{sec:model}, we introduce the mixed spin-$1/2$ and spin-$1$ Ising model on the Lieb lattice and outline its exact solution based on the generalized decoration-iteration transformation. Section~\ref{gs} presents the ground-state properties and analyzes the effective mapping parameters governing the existence of finite-temperature phase transitions. In Sec.~\ref{pt}, we determine the complete finite-temperature phase diagram with particular emphasis on the interplay between the magnetic field and the uniaxial single-ion anisotropy leading to discontinuous, continuous, and reentrant thermal phase transitions. The exact analytical predictions are subsequently verified by classical Monte Carlo simulations in Sec.~\ref{mc}. In Sec.~\ref{sim}, we compare the obtained results with those previously reported for the spin-$1/2$ Ising-Heisenberg model on the diamond-decorated square lattice and discuss their common physical origin. Finally, the main findings are summarized in Sec.~\ref{conc}.

\section{Model and methods}
\label{sec:model}

\begin{figure}[t]
\centering
\includegraphics[width=0.4\linewidth]{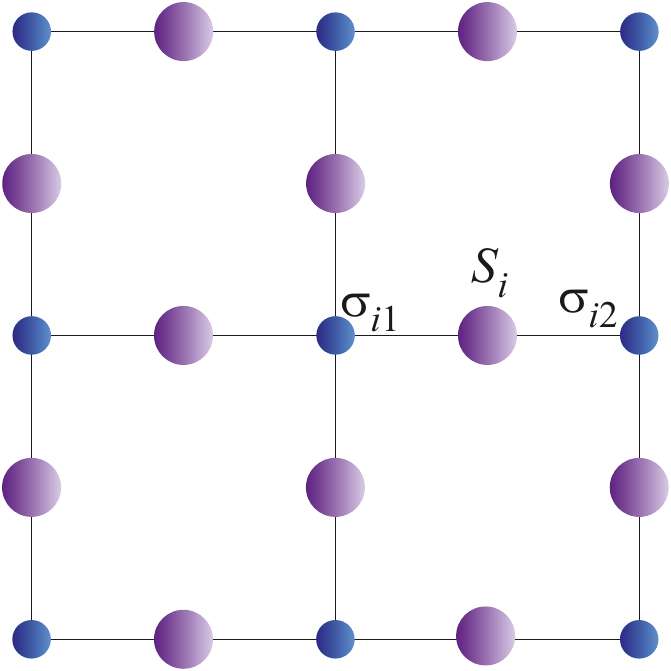}
\caption{Schematic illustration of the mixed spin-$1/2$ and spin-$1$ Ising model on a Lieb lattice. The nodal sites (small blue circles) are occupied by the spin-$1/2$ variables $\sigma_i=\pm1/2$, whereas the decorating sites (large violet circles) are occupied by the spin-$1$ variables $S_i=0,\pm1$. Each decorating spin $S_i$ on the $i$-th bond interacts with two adjacent nodal spins $\sigma_{i1}$ and $\sigma_{i2}$.}
\label{fig:lattice}
\end{figure}

We consider a mixed spin-$1/2$ and spin-$1$ Ising model on a Lieb lattice schematically illustrated in Fig.~\ref{fig:lattice}. The nodal sites are occupied by the spin-$1/2$ variables $\sigma_i=\pm 1/2$, whereas the decorating sites are occupied by the spin-$1$ variables $S_i=0,\pm1$. The Hamiltonian of the mixed spin-$1/2$ and spin-$1$ Ising model on the Lieb lattice in a magnetic field is then given by
\begin{equation}
\mathcal{H}= J \sum_{\langle i, j \rangle} S_i \sigma_j 
+ D \sum_{i=1}^{2N} S_i^2 - h \sum_{i=1}^{2N} S_i - h \sum_{j=1}^{N} \sigma_j,
\label{eq:ham}
\end{equation}
 where $J$ denotes the nearest-neighbor interaction between decorating and nodal spins, $D$ is the uniaxial single-ion anisotropy acting on the decorating spin-$1$ variables, the last two terms represent the Zeeman energy of the decorating and nodal spins in the external magnetic field $h$, and $N$ denotes the total number of the nodal spins. The total Hamiltonian (\ref{eq:ham}) can be decomposed into a sum of bond Hamiltonians
\begin{equation}
\mathcal{H}=\sum_{i=1}^{2N}\mathcal{H}_i,
\label{eq:hamiltonian_total}
\end{equation}
where each bond Hamiltonian $\mathcal{H}_i$ contains all interaction terms associated with the decorating spin $S_i$ located on the $i$-th bond 
\begin{equation}
\mathcal H_i = J S_i(\sigma_{i1}+\sigma_{i2}) + D S_i^2 - h S_i - \frac{h}{4}(\sigma_{i1}+\sigma_{i2}).
\label{eq:bond_hamiltonian}
\end{equation}
The factor $1/4$ in the Zeeman term of the nodal spins prevents its overcounting, since this term is distributed equally among the four bond Hamiltonians associated with its nearest-neighbor
decorating spins.

The partition function of the mixed-spin Ising model on a Lieb lattice can be partially factorized into the following product
\begin{equation}
{\cal Z}=\sum_{\{\sigma\}}\prod_{i=1}^{2N} \sum_{S_i=0,\pm1} \!\!\! 
\exp\left(-\beta\mathcal{H}_i\right),
\label{eq:partition_factorized}
\end{equation}
where the summation $\sum_{\{\sigma\}}$ extends over all configurations of the nodal spins, while the summations over states of the decorating spins can be carried out independently of each other. After performing the summation over the three possible states of the decorating spin $S_i=0,\pm1$ one obtains the resulting Boltzmann weight, which can be subsequently replaced through the generalized decoration-iteration transformation \cite{Fisher1959,Syozi1972,Rojas2009}
\begin{eqnarray}
\sum_{S_i=0,\pm1} \!\!\! \exp\left(-\beta\mathcal{H}_i\right)
\!\!\!\!&=&\!\!\!\! 
\exp\Bigl[\frac{\beta h}{4}(\sigma_{i1}\!+\!\sigma_{i2})\Bigr]
\left\{
1\!+\!2 \exp(-\beta D) \cosh \left[\beta J(\sigma_{i1}\!+\!\sigma_{i2})\!-\!\beta h\right]
\right\} \nonumber \\
\!\!\!\!&=&\!\!\!\! A \exp \Bigl[\beta J_{\rm eff}\sigma_{i1}\sigma_{i2} + \frac{\beta h_{\rm eff}}4 (\sigma_{i1}+\sigma_{i2}) \Bigr].
\label{eq:DIT}
\end{eqnarray}
The mapping parameters $A$, $J_{\rm eff}$, and $h_{\rm eff}$ are determined from the self-consistency condition requiring that the decoration-iteration transformation~(\ref{eq:DIT}) reproduces three independent expressions obtained from Eq. (\ref{eq:DIT}) by considering four available states of the nodal spins $\sigma_{i1}$ and $\sigma_{i2}$
\begin{align}
A &= (V_1V_2V_3^2)^{1/4}, \label{eq:A} \\
\beta J_{\rm eff} &= \ln\left( \frac{V_1V_2}{V_3^2} \right), \label{eq:Jeff} \\
\beta h_{\rm eff} &= \beta h + 2\ln\left(\frac{V_1}{V_2}\right), \label{eq:heff}
\end{align}
which are defined through the following functions
\begin{align}
V_1 &= 1 + 2 \exp(-\beta D) \cosh(\beta J - \beta h), \\
V_2 &= 1 + 2 \exp(-\beta D) \cosh(\beta J + \beta h), \\
V_3 &= 1 + 2 \exp(-\beta D) \cosh(\beta h).
\label{eq:v123}
\end{align}
The mapping parameter $A$ represents a simpler multiplicative factor, whereas the mapping parameters $J_{\rm eff}$ and $h_{\rm eff}$ determine the effective interaction between the nodal spins and the effective field acting on them, respectively. By substituting the decoration-iteration transformation (\ref{eq:DIT}) into the factorized form of the partition function (\ref{eq:partition_factorized}), one establishes an exact mapping correspondence between the partition function of the mixed spin-$1/2$ and spin-$1$ Ising model on a Lieb lattice and the partition function of an effective spin-$1/2$ Ising model on the square lattice 
\begin{equation}
{\cal Z} (\beta, J, D, h) = A^{2N} {\cal Z}_{\rm eff} (\beta, J_{\rm eff}, h_{\rm eff}),
\label{eq:partition_mapping}
\end{equation}
which corresponds to the following effective Hamiltonian
\begin{equation}
{\cal H}_{\rm eff} = - J_{\rm eff} \sum_{\langle i,j \rangle} \sigma_{i} \sigma_{j}
 - h_{\rm eff} \sum_{i=1}^N \sigma_{i}.
\label{eq:heff}
\end{equation}
The exact mapping relation (\ref{eq:partition_mapping}) between the two partition functions enables one to derive all magnetic and thermodynamic quantities of the investigated mixed-spin Ising model on the Lieb lattice from the corresponding quantities of the effective spin-$1/2$ Ising model on the square lattice. For instance, the Gibbs free energy of the mixed spin-1/2 and spin-1 Ising model on the Lieb lattice can be expressed in terms of the Gibbs free energy of the effective spin-$1/2$ Ising model on the square lattice as
\begin{equation}
G = G_{\rm eff} - 2N k_{\rm B} T \ln A. 
\label{eq:gibbs_free_energy}
\end{equation}

In addition, the exact mapping theorems \cite{Barry1988,Barry1990,Barry1991,Barry1995} state that any statistical average depending solely on the nodal spins remains invariant under the mapping transformation
\begin{equation}
\langle f(\sigma_i, \sigma_j, \ldots, \sigma_k) \rangle = 
\langle f(\sigma_i, \sigma_j, \ldots, \sigma_k) \rangle_{\rm eff}, 
\end{equation}
where $\langle \cdots \rangle$ and $\langle \cdots \rangle_{\rm eff}$ denote canonical ensemble averages evaluated in the original and corresponding effective models, respectively. The single-site magnetization of the nodal spins can be accordingly computed from the relation  
\begin{equation}
m_\sigma \equiv \langle \sigma_i\rangle = \langle \sigma_i\rangle_{\rm eff} \equiv
m_{\rm eff}(\beta,J_{\rm eff},h_{\rm eff}).
\label{eq:m_sigma_definition}
\end{equation}
Similarly, the nearest-neighbor pair correlation function follows from the formula
\begin{equation}
\varepsilon_\sigma \equiv \langle \sigma_{i1}\sigma_{i2}\rangle
= \langle \sigma_{i1}\sigma_{i2}\rangle_{\rm eff} \equiv
\varepsilon_{\rm eff}(\beta,J_{\rm eff},h_{\rm eff}).
\label{eq:epsilon_sigma_definition}
\end{equation}
On the other hand, the single-site magnetization of the decorating spins can be calculated by exploiting the exact Callen-Suzuki identity \cite{Callen63,Suzuki65,Balcerzak02} 
\begin{equation}
m_S \equiv \langle S_i \rangle = \left\langle
\frac{\sum_{S_i=0,\pm1} S_i \exp(-\beta\mathcal{H}_i)}{\sum_{S_i=0,\pm1}\exp(-\beta\mathcal{H}_i)}
\right\rangle
= \left\langle \frac{-2 \sinh \left[\beta J(\sigma_{i1}\!+\!\sigma_{i2})\!-\!\beta h\right]}{\exp(\beta D) + 2 \cosh \left[\beta J(\sigma_{i1}\!+\!\sigma_{i2})\!-\!\beta h\right]} \right\rangle.
\label{eq:m_S_definition}
\end{equation}
To calculate statistical average appearing in Eq. (\ref{eq:m_S_definition}), one may first expand the relevant expression in terms of two nodal spins $\sigma_{i1}$ and $\sigma_{i2}$ involved therein
\begin{equation}
F (\sigma_{i1}, \sigma_{i2}) = \frac{-2 \sinh \left[\beta J(\sigma_{i1}\!+\!\sigma_{i2})\!-\!\beta h\right]}{\exp(\beta D) + 2 \cosh \left[\beta J(\sigma_{i1}\!+\!\sigma_{i2})\!-\!\beta h\right]}
= A_0 + A_1 \sigma_{i1} + A_2 \sigma_{i2} + A_{12}\sigma_{i1}\sigma_{i2}.
\label{eq:f_expansion}
\end{equation}
Considering the four possible configurations of the two nodal spins $\sigma_{i1}$ and $\sigma_{i2}$ it is convenient to introduce three auxiliary functions derived from the expression $F(\sigma_{i1},\sigma_{i2})$
\begin{align}
F_1 &\equiv -F\left(+\frac{1}{2}, +\frac{1}{2}\right) =
\frac{2\sinh(\beta J-\beta h)}{\exp(\beta D)+2\cosh(\beta J-\beta h)}, \label{eq:F1} \\
F_2 &\equiv +F\left(-\frac{1}{2}, -\frac{1}{2}\right) =
\frac{2\sinh(\beta J+\beta h)}{\exp(\beta D)+2\cosh(\beta J+\beta h)}, \label{eq:F2} \\
F_3 &\equiv +F\left(\pm \frac{1}{2}, \mp \frac{1}{2}\right) =
\frac{2\sinh(\beta h)}{\exp(\beta D)+2\cosh(\beta h)}, \label{eq:F3}
\end{align}
which in turn unambiguously determine unknown expansion coefficients 
\begin{align}
A_0 &= -\frac{1}{4} \left(F_1 - F_2 - 2F_3\right), \label{eq:A0} \\
A_1 &= A_2 = -\frac{1}{2} \left(F_1 + F_2\right), \label{eq:A1} \\
A_{12} &= -(F_1 - F_2 + 2F_3). \label{eq:A12} 
\end{align} 
By substituting the expansion coefficients (\ref{eq:A0})-(\ref{eq:A12}) into Eq.~(\ref{eq:f_expansion}) and performing the corresponding statistical average, one obtains the final expression for the single-site magnetization of the decorating spins
\begin{equation}
m_S = -\frac14(F_1-F_2-2F_3) - m_\sigma(F_1+F_2) - \varepsilon_\sigma(F_1-F_2+2F_3), 
\label{eq:m_S_final}
\end{equation}
which is expressed in terms of the single-site magnetization of the nodal spins $m_\sigma$ and the pair correlation function between the nearest-neighbor nodal spins $\varepsilon_\sigma$ given by Eqs. (\ref{eq:m_sigma_definition}) and (\ref{eq:epsilon_sigma_definition}), respectively. 

The exact mapping relation derived above becomes particularly useful when the effective field vanishes. Under this condition, the investigated mixed-spin Ising model on the Lieb lattice in a magnetic field is rigorously mapped onto the exactly solved spin-$1/2$ Ising model on the square lattice in zero field \cite{Onsager1944}. Hence, the condition of exact solvability is limited to the particular parameter subspace with zero effective field $h_{\rm eff}=0$, which is according to Eq. (\ref{eq:heff}) equivalent with the condition
\begin{equation}
\exp \left(\frac{\beta h}{2}\right) = 
\frac{1 + 2 \exp(-\beta D) \cosh(\beta J+\beta h)}{1 + 2 \exp(-\beta D) \cosh(\beta J-\beta h)}. 
\label{eq:heff_zero}
\end{equation}
Although the condition of zero effective field (\ref{eq:heff_zero}) is trivially satisfied for arbitrary temperature in the absence of a genuine external magnetic field $h=0$, this condition also possesses a nontrivial solution for certain temperatures even at finite genuine magnetic fields $h \neq 0$ (see below). Within the particular manifold of the parameter space with zero effective field (\ref{eq:heff_zero}), the system undergoes a continuous phase transition when the effective interaction reaches the critical value $\beta_c J_{\rm eff} = 2\ln(1+\sqrt2)$ that consequently also yields the exact critical condition for the mixed spin-$1/2$ and spin-$1$ Ising model on the Lieb lattice in a magnetic field
\begin{eqnarray}
\frac{1 + 2 \exp(-\beta_c D) \cosh(\beta_c J + \beta_c h)}{1 + 2 \exp(-\beta_c D) \cosh(\beta_c h)} = (1+\sqrt2) \exp \left( \frac{\beta_c h}{4} \right), 
\label{eq:critical_line}
\end{eqnarray}
where $\beta_c = 1/(k_{\rm B} T_c)$ and $T_c$ is the critical temperature. Furthermore, discontinuous thermal phase transitions are expected to occur in the mixed spin-$1/2$ and spin-1 Ising model on a Lieb lattice within the parameter subspace (\ref{eq:heff_zero}) satisfying the condition of vanishing effective field $h_{\rm eff} = 0$ provided that the temperature is set below its critical value, i.e. $\beta J_{\rm eff} > 2 \ln (1+\sqrt{2})$.  

\section{Ground-state properties}
\label{gs}

\begin{figure}[t]
\centering
\includegraphics[width=0.55\linewidth]{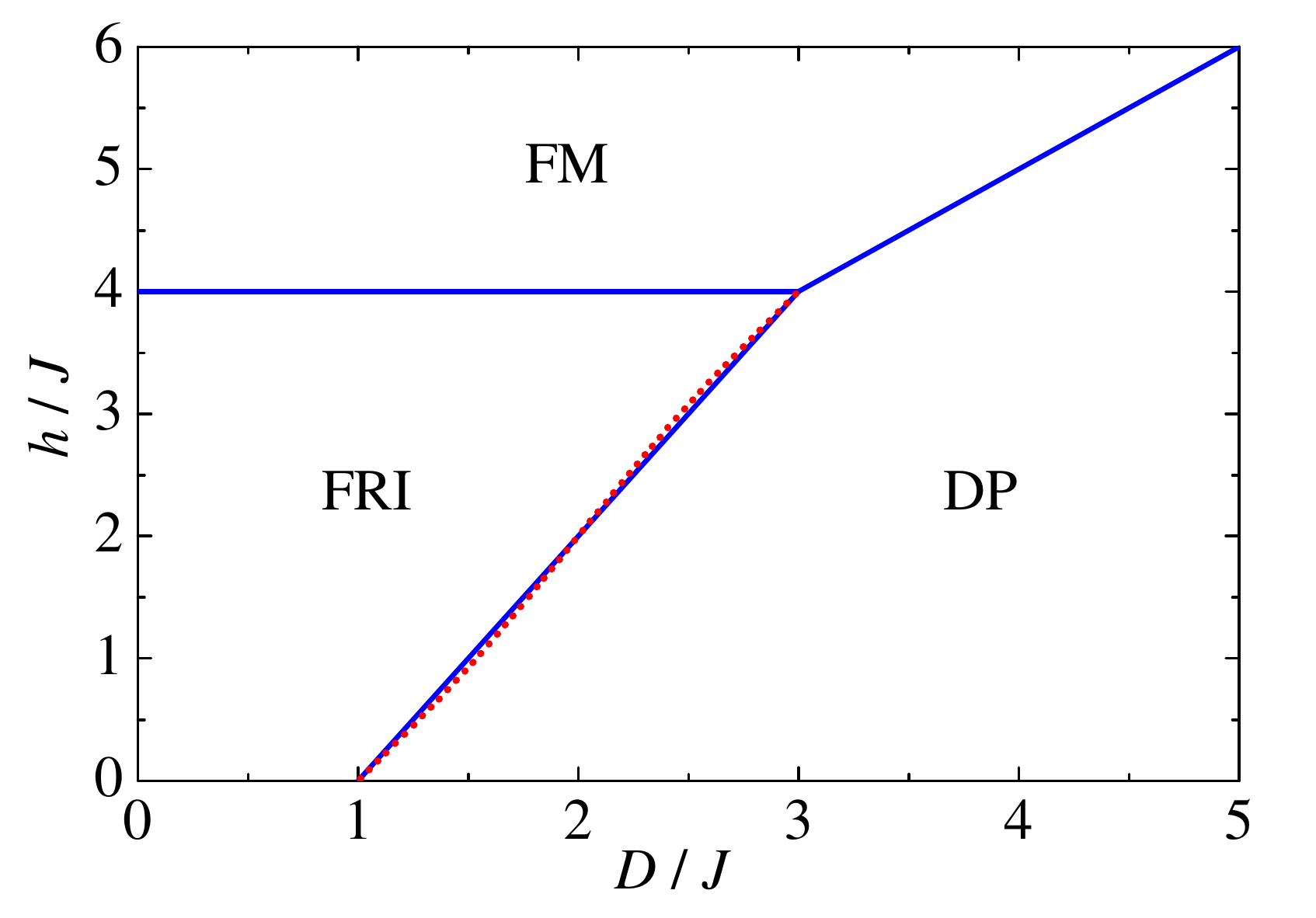}
\caption{Ground-state phase diagram of the mixed spin-$1/2$ and spin-$1$ Ising model on the Lieb lattice in the $D/J$-$h/J$ plane. Three distinct phases are identified: the ferrimagnetic phase (FRI), the disordered phase (DP), and the ferromagnetic phase (FM). The solid blue lines indicate the phase boundaries obtained by comparing the corresponding ground-state energies, whereas the red dashed line closely following the coexistence line between the FRI and DP phases represents the projection of the finite-temperature critical line (\ref{eq:critical_line}) into a zero-temperature plane.}
\label{gspd1}
\end{figure}

The ground-state phase diagram of the mixed spin-$1/2$ and spin-$1$ Ising model on the Lieb lattice is depicted in Fig.~\ref{gspd1} in the $D/J$-$h/J$ plane. Depending on a relative strength of the uniaixial single-ion anisotropy $D/J$ and the external magnetic field $h/J$, three different ground states separated by lines of discontinuous phase transitions can be identified. The ferrimagnetic phase (FRI) is characterized by an antiparallel spin alignment of the decorating and nodal spins acquiring the values $S_i=1$ and $\sigma_i=-1/2$, respectively. The nonmagnetic spin state $S_i=0$ of the decorating spins is favored by the easy-plane uniaxial single-ion anisotropy in the disordered phase (DP), which additionally involves the nodal spins $\sigma_i=1/2$ fully polarized along the magnetic-field direction. Finally, the ferromagnetic phase (FM) is characterized by a full alignment of all decorating and nodal spins into the external magnetic field, i.e., $S_i=1$ and $\sigma_i=1/2$. The phase boundaries separating these three phases are readily determined by comparing  their respective ground-state energies. Specifically, the coexistence line between the FRI and DP phases is given by the formula $h/J=2(D/J-1)$, whereas the FRI-FM and DP-FM phase boundaries follow from the conditions $h/J=4$ and $h/J=1+D/J$, respectively. It is worth mentioning that all three phase boundaries meet at the triple point with coordinates $[D/J;h/J]=[3;4]$, where the FRI, DP, and FM phases coexist together. 

\begin{figure}[t]
\centering
\includegraphics[width=0.49\linewidth]{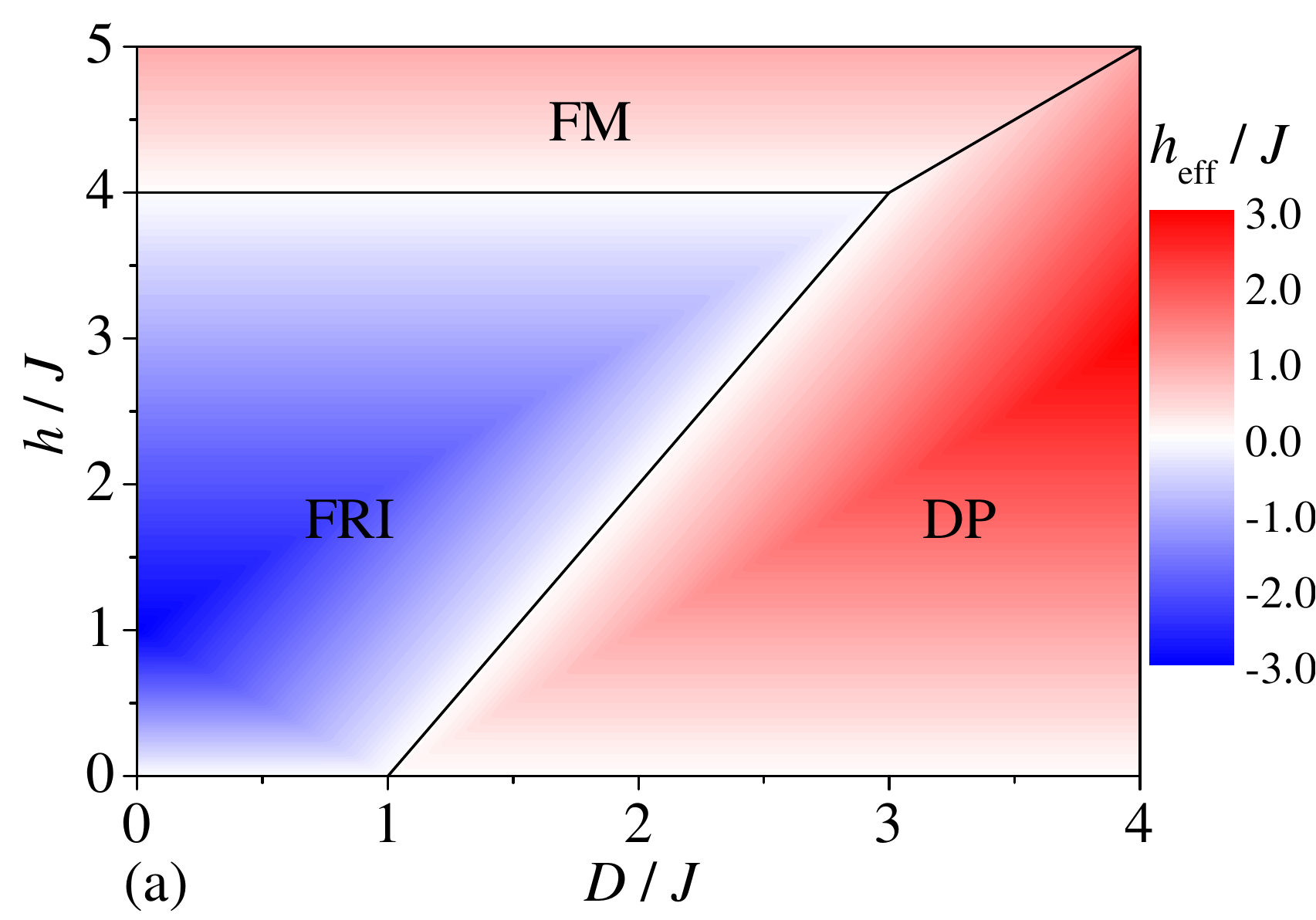}
\includegraphics[width=0.49\linewidth]{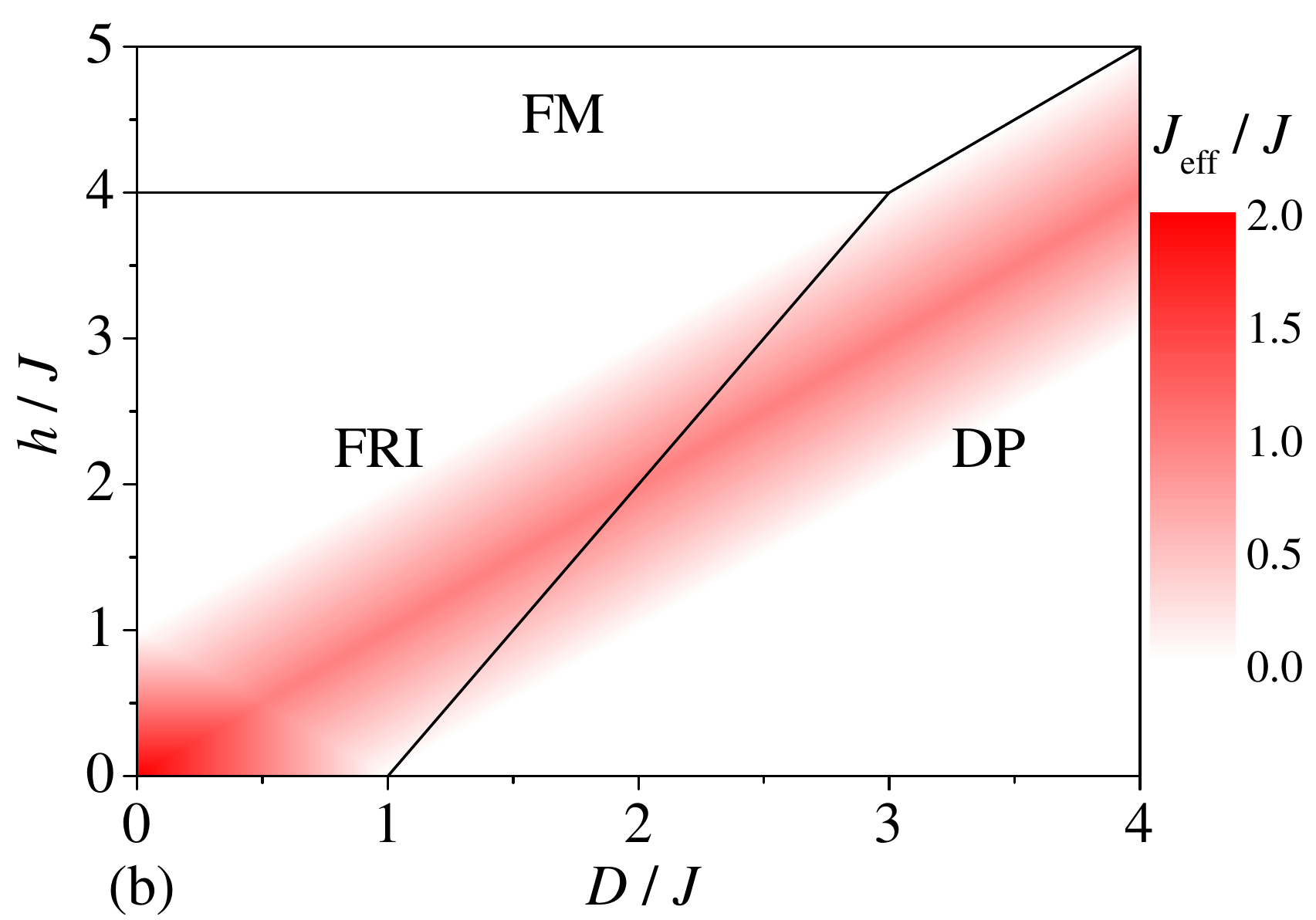}
\caption{Zero-temperature density plots of the effective magnetic field $h_{\rm eff}/J$ (a) and the effective interaction $J_{\rm eff}/J$ (b) in the $D/J-h/J$ plane. The solid black lines represent ground-state phase boundaries separating the ferrimagnetic (FRI), disordered (DP), and ferromagnetic (FM) phases.}
\label{fig:hJeff}
\end{figure}

Figure \ref{fig:hJeff} illustrates zero-temperature asymptotic values of the effective field $h_{\rm eff}$ and effective interaction $J_{\rm eff}$ resulting from a mapping correspondence with the effective spin-1/2 Ising square lattice established via the generalized decoration-iteration transformation (\ref{eq:DIT}). As shown in Fig.~\ref{fig:hJeff}(a), the effective field changes sign upon crossing the ground-state phase boundaries separating the FRI phase from the other two phases. The effective field is negative $h_{\rm eff}<0$ throughout the FRI phase, whereas it becomes positive 
$h_{\rm eff}>0$ in both the DP and FM phases. Hence, the zero-field condition $h_{\rm eff}=0$ is satisfied along both FRI-DP and FRI-FM coexistence lines.

A markedly different behavior is observed for the effective interaction shown in Fig.~\ref{fig:hJeff}(b). First, the effective interaction remains non-negative throughout the entire parameter region implying that the original model is always mapped onto a ferromagnetic spin-$1/2$ Ising model on the square lattice. Moreover, the effective interaction vanishes along the FRI-FM and DP-FM coexistence lines, whereas it remains finite along the FRI-DP phase boundary and reaches its maximum value along the line $h/J=D/J$. It is noteworthy that a nonzero value of the effective interaction represents a necessary prerequisite for the occurrence of a phase transition.

The combined behavior of the effective field and effective interaction consequently provides valuable insight into the possible existence of finite-temperature phase transitions. The effective ferromagnetic spin-1/2 Ising model on the square lattice undergoes a phase transitions only when a non-vanishing effective interaction $J_{\rm eff}>0$ is accompanied by strictly zero effective field $h_{\rm eff}=0$. Since these two conditions are simultaneously fulfilled only along the FRI-DP coexistence line, this ground-state phase boundary represents the sole candidate for the emergence of finite-temperature phase transitions and criticality. This prediction will be examined in detail in the following section.

\section{Thermal phase transitions}
\label{pt}

\begin{figure}[t]
\centering
\includegraphics[width=0.75\linewidth]{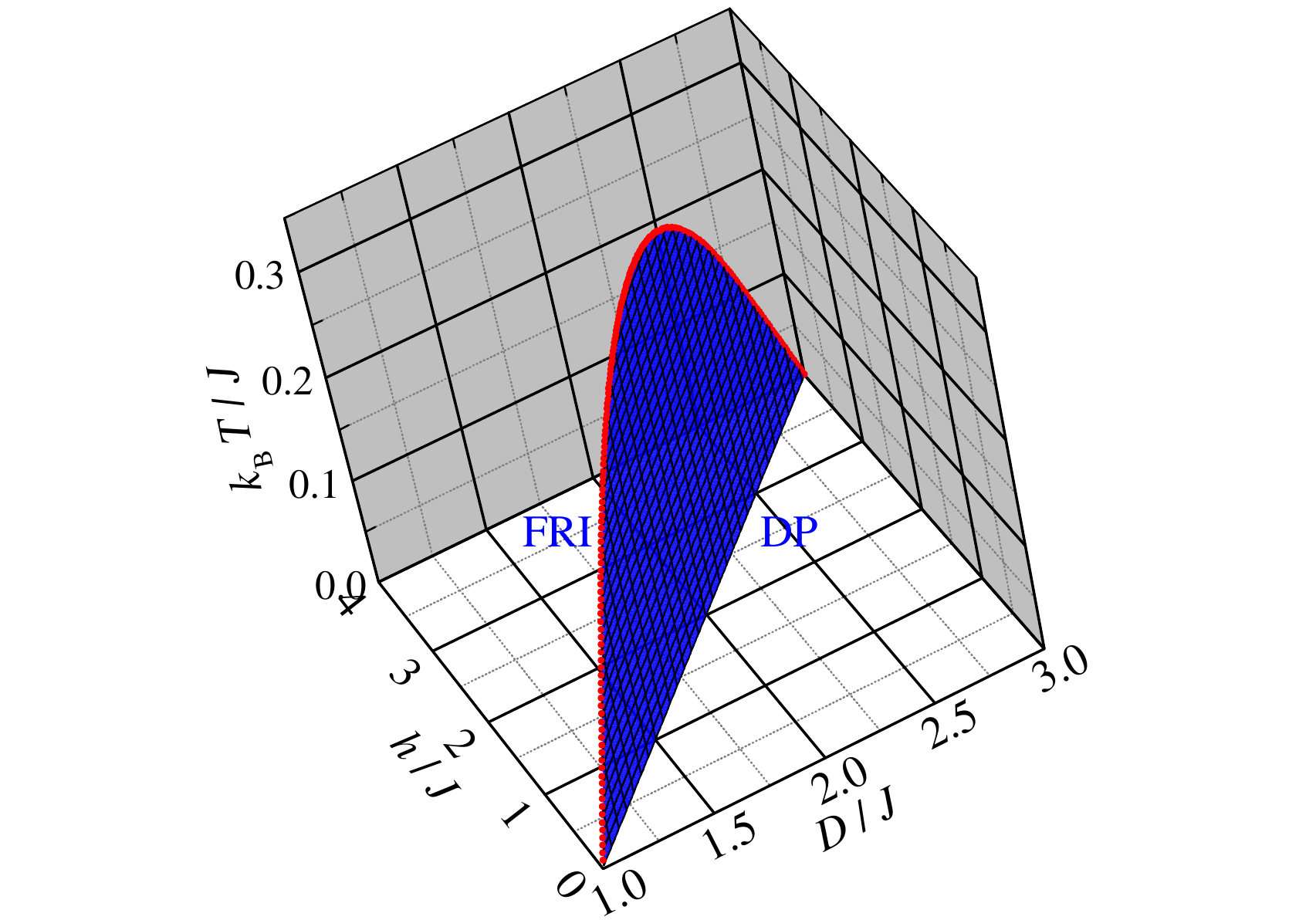}
\caption{Global phase diagram of the mixed spin-$1/2$ and spin-$1$ Ising model on the Lieb lattice in the three-dimensional parameter space $D/J-h/J-k_{\rm B}T/J$. The blue dome-shaped surface represents the locus of discontinuous phase transitions separating the ferrimagnetic (FRI) and disordered (DP) phases. A red critical line corresponds to continuous phase transitions between these phases.}
\label{fig:plocha}
\end{figure}

Finite-temperature phase transitions and critical behavior of the mixed spin-$1/2$ and spin-$1$ Ising model on the Lieb lattice are summarized in Fig.~\ref{fig:plocha} in the form of three-dimensional plot in the parameter space $D/J-h/J-k_{\rm B}T/J$. The global phase diagram contains a dome-shaped surface of discontinuous phase transitions exactly determined by the condition of zero effective field (\ref{eq:heff_zero}), which is bounded from above by a critical line of continuous phase transitions given exactly by the critical condition (\ref{eq:critical_line}). The dome-shaped wall apparently separates the FRI phase from the DP phase as it emerges from their common ground-state phase boundary. Consistently, the projection of the critical line (\ref{eq:critical_line}) onto the zero-temperature plane also closely follows this FRI--DP coexistence line (see Fig. \ref{gspd1}). This finding thus fully confirms the predictions based on the zero-temperature behavior of the effective interaction and effective field. The critical temperature exhibits a highly nonmonotonic dependence on both the uniaxial single-ion anisotropy and the external magnetic field when starting from zero at the onset $[D/J;h/J]=[1;0]$ of the FRI-DP ground-state phase boundary, then increasing continuously to a pronounced maximum located approximately at its midpoint $D/J=2$, and subsequently decreasing again until it completely vanishes at its end point, which coincides with the triple point $[D/J;h/J]=[3;4]$ where all three FRI, DP and FM phases coexist.

\begin{figure}[!ht]
\centering
\includegraphics[width=0.49\linewidth]{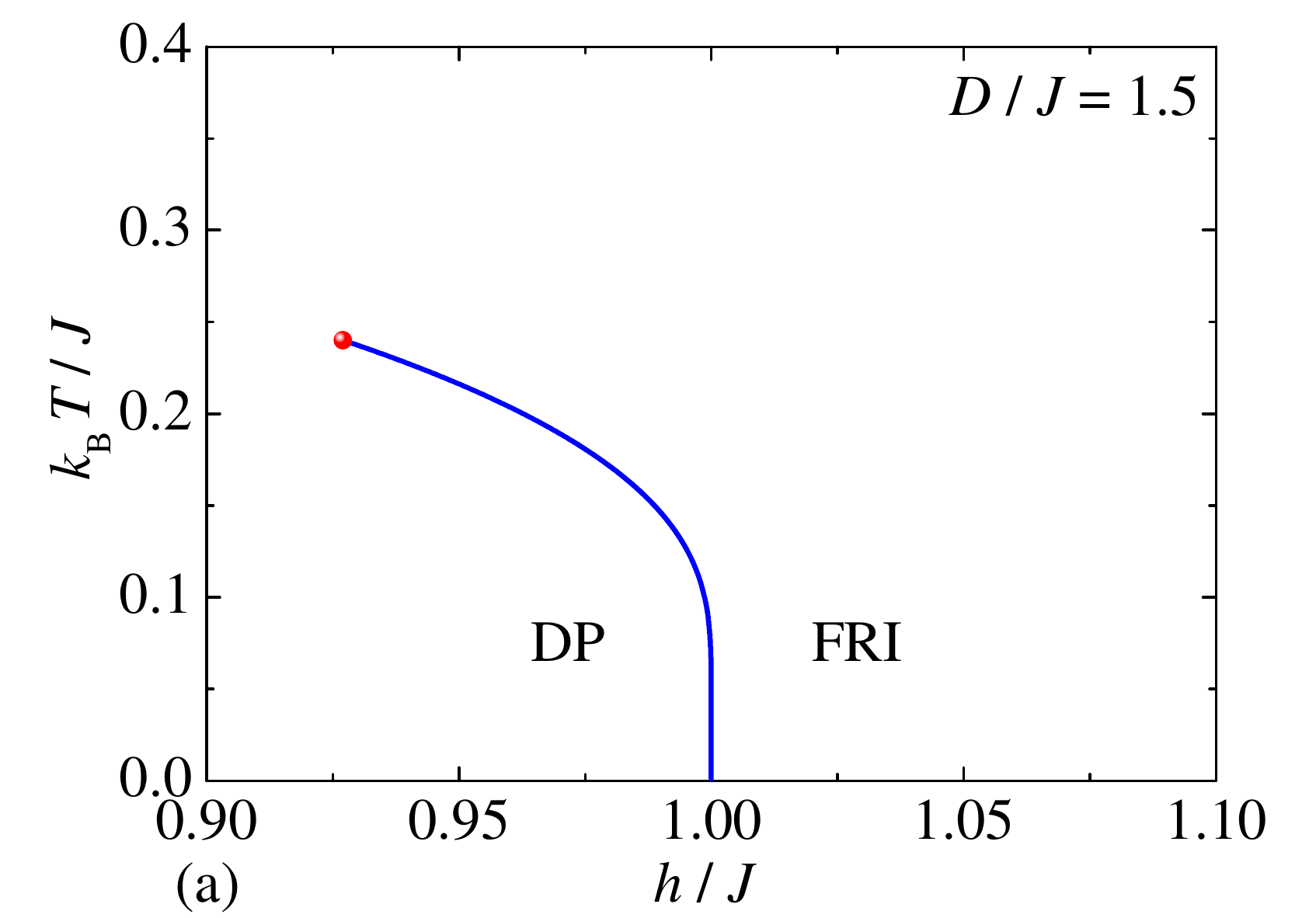}
\includegraphics[width=0.49\linewidth]{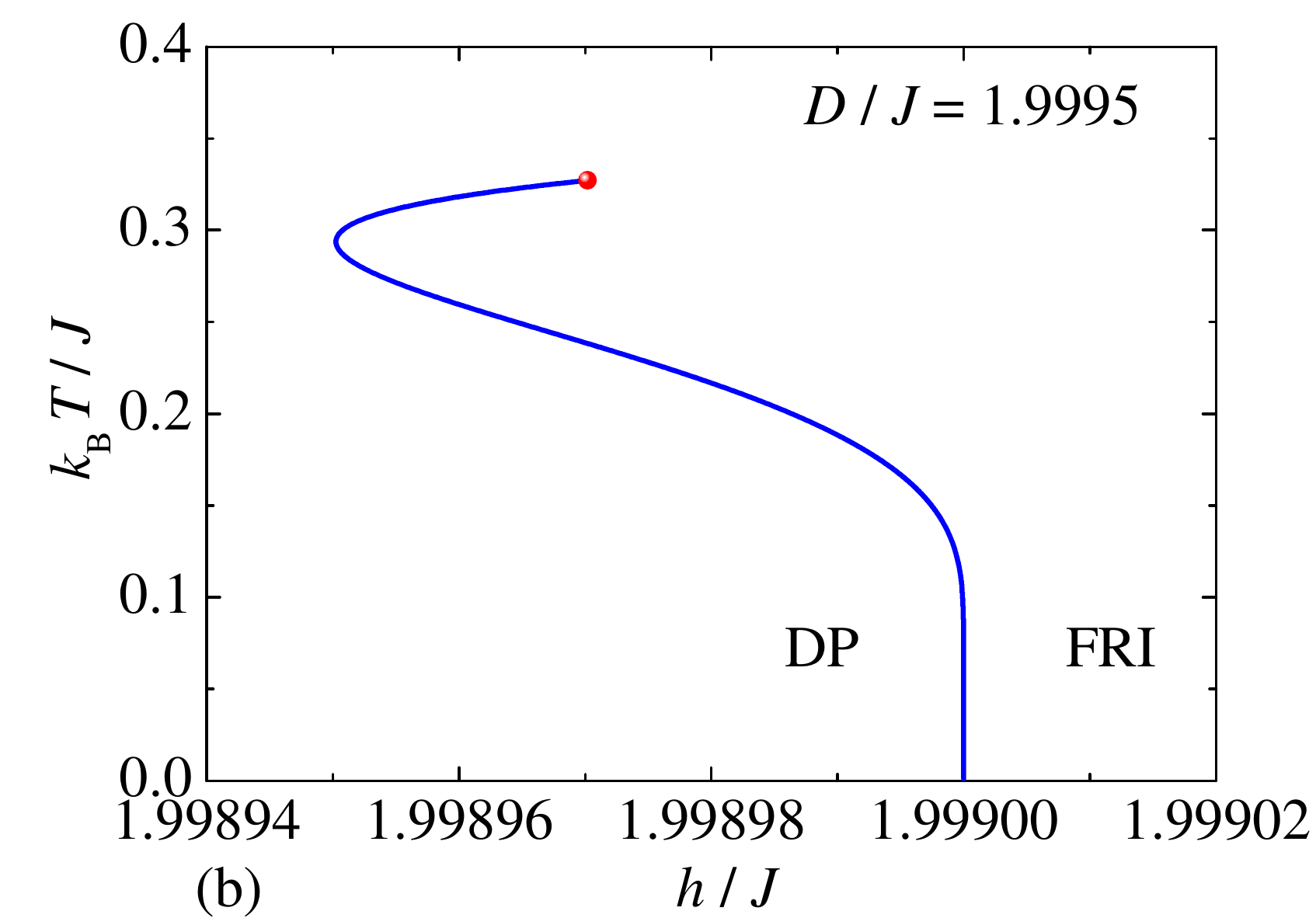}
\includegraphics[width=0.49\linewidth]{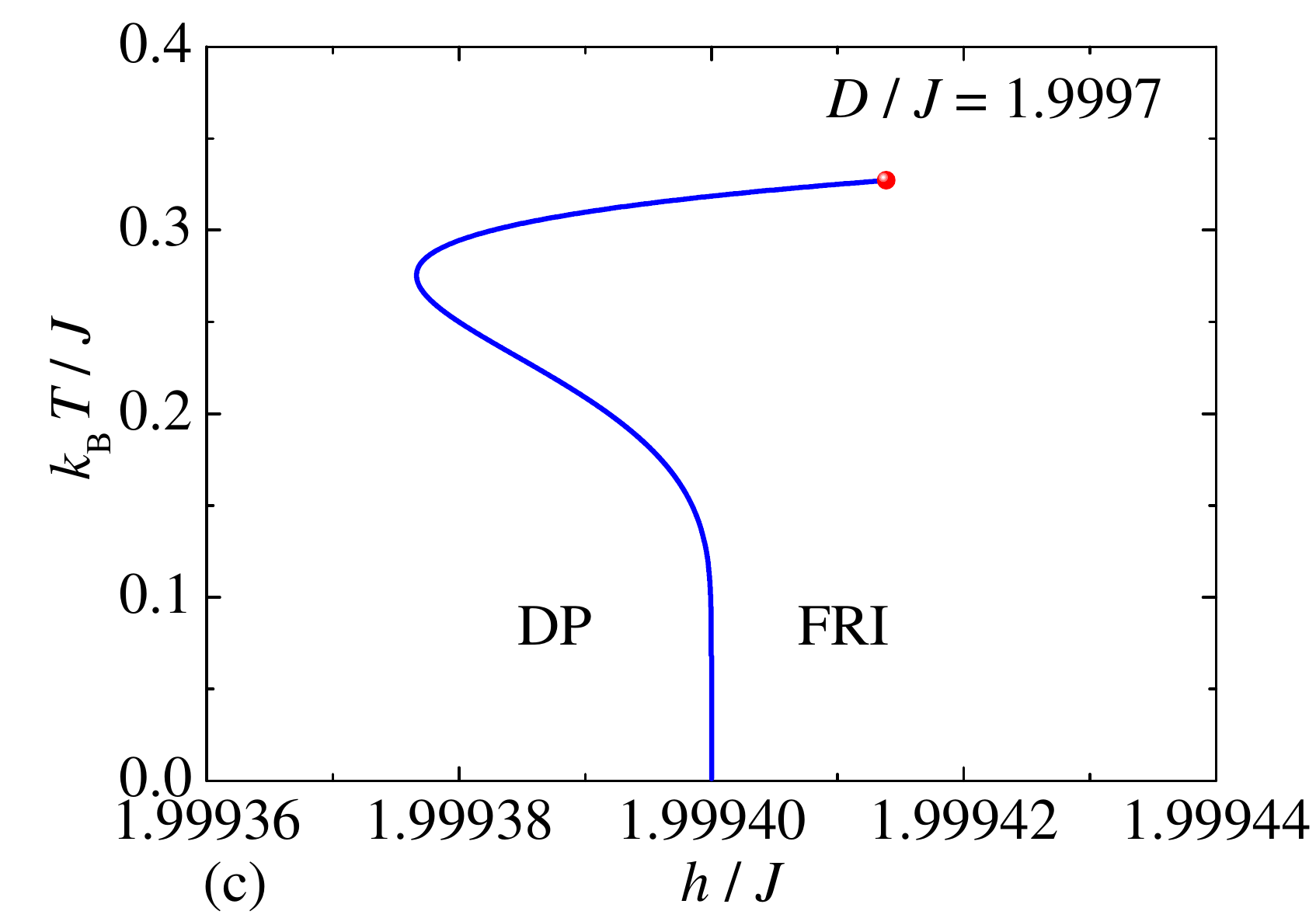}
\includegraphics[width=0.49\linewidth]{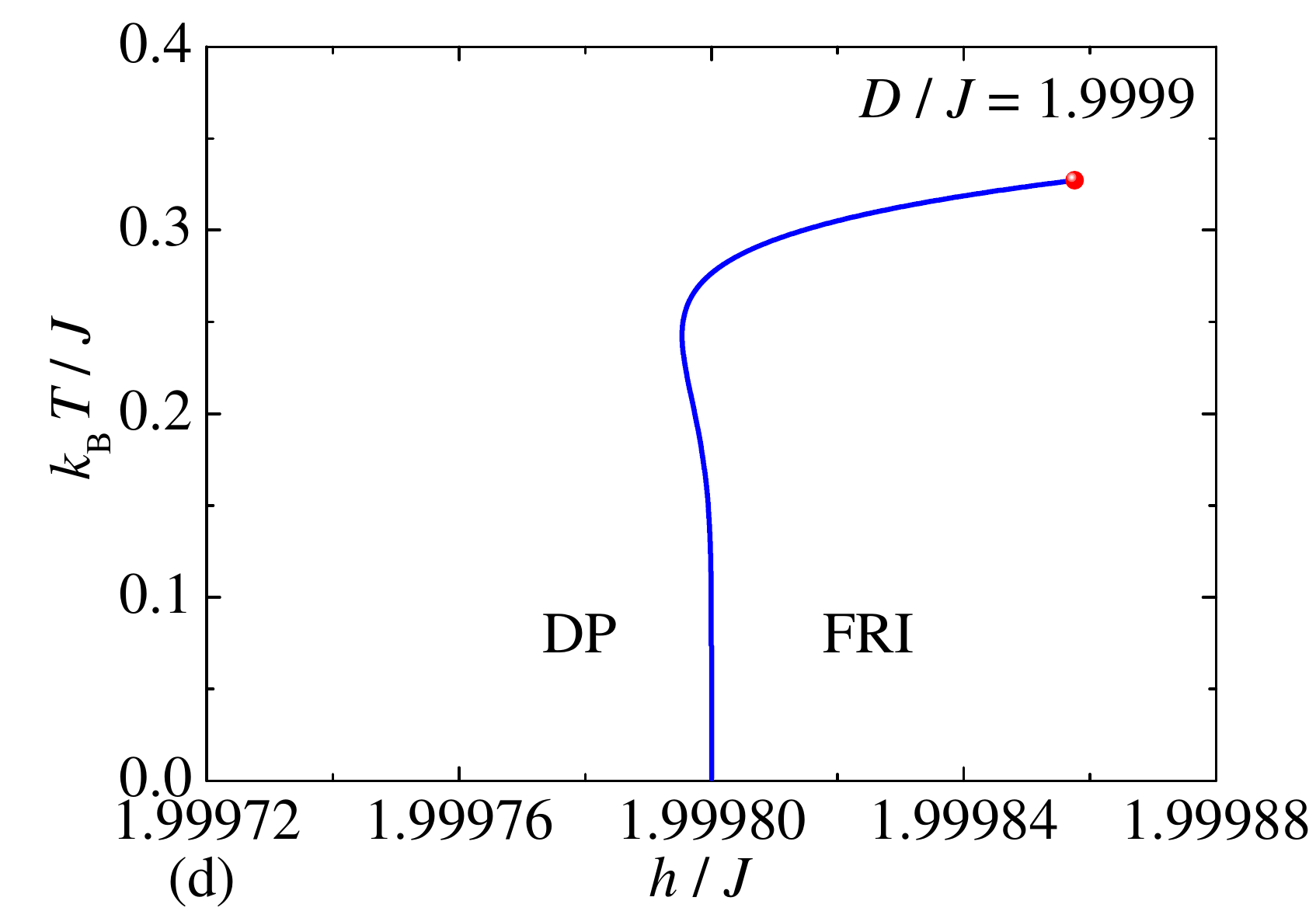}
\includegraphics[width=0.49\linewidth]{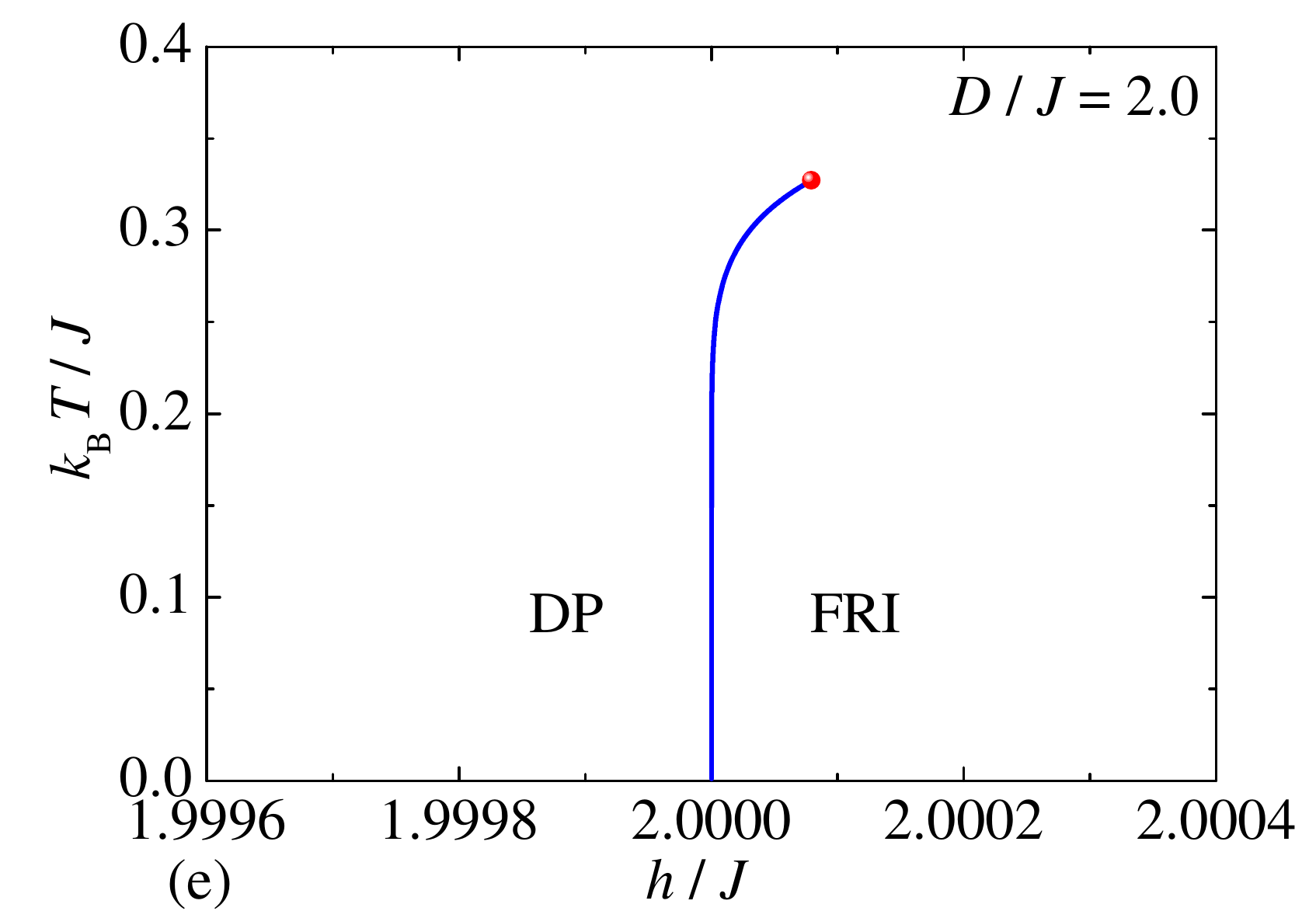}
\includegraphics[width=0.49\linewidth]{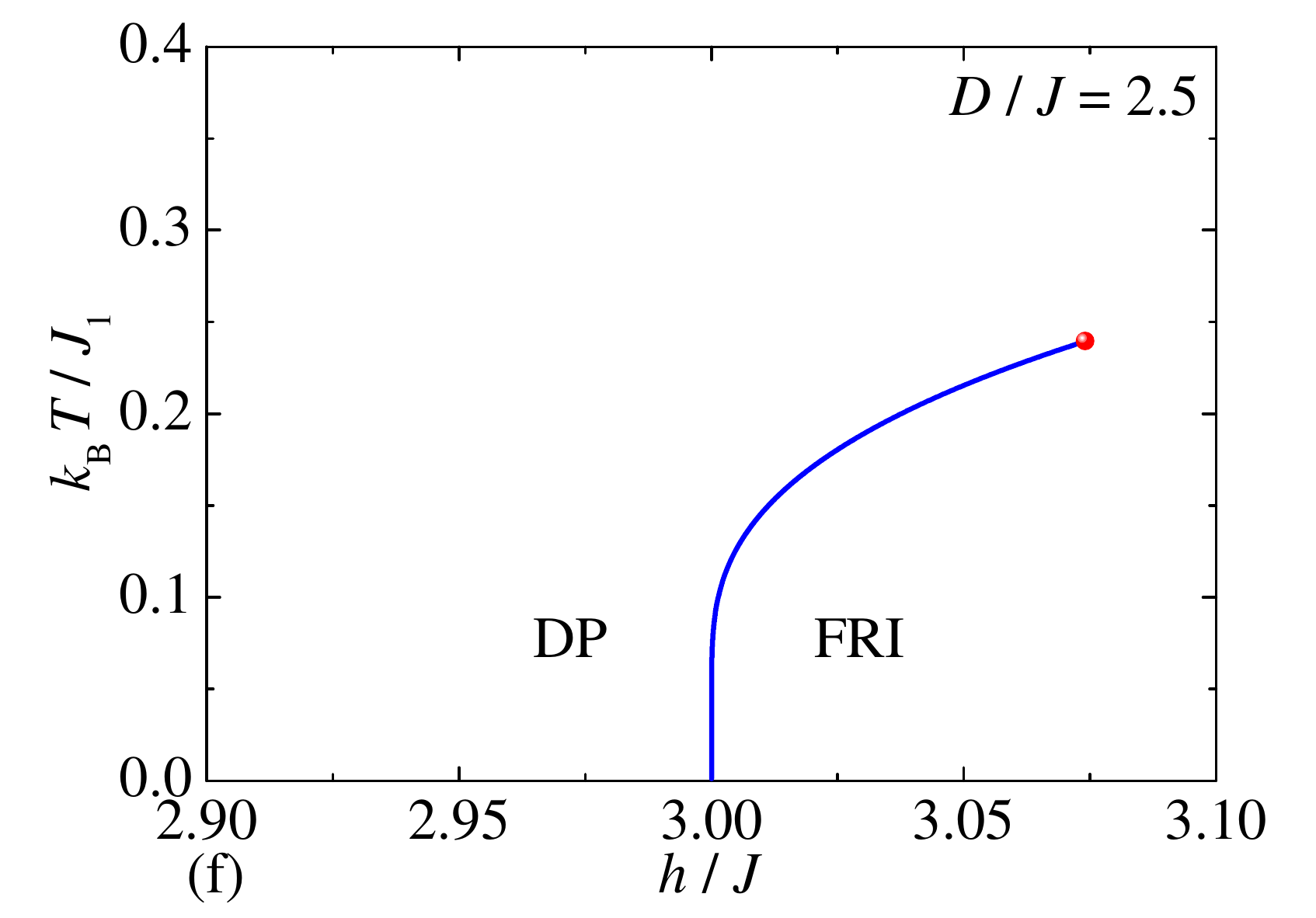}
\caption{Representative magnetic-field dependencies of the transition temperature of the mixed spin-$1/2$ and spin-$1$ Ising model on the Lieb lattice calculated for six selected values of the uniaxial single-ion anisotropy: (a) $D/J=1.5$, (b) $D/J=1.9995$, (c) $D/J=1.9997$, (d) $D/J=1.9999$, (e) $D/J=2.0$, and (f) $D/J=2.5$. The lines of discontinuous phase transitions separating the ferrimagnetically ordered phase (FRI) from the disordered phase (DP) terminate at Ising-type critical points (red circles).}
\label{fig:reentrant}
\end{figure}

To gain deeper insight into the evolution of thermal phase transitions, Fig.~\ref{fig:reentrant} displays representative magnetic-field dependencies of the transition temperature for several selected values of the uniaxial single-ion anisotropy. The displayed lines of discontinuous thermal phase transitions separating the FRI phase from the DP phase terminate at the critical points, where the thermal phase transition changes its character from discontinuous to continuous. These terminal critical points belong to the universality class of the two-dimensional Ising model in accordance with the exact mapping correspondence established in Sec.~\ref{sec:model}. For $D/J=1.5$, the transition line exhibits a conventional monotonic temperature dependence when originating from a coexistence point of FRI and DP phases, then bending toward lower magnetic fields before terminating at a critical point [Fig.~\ref{fig:reentrant}(a)]. The most remarkable profile of the transition lines can be observed when the magnetic field is fixed to $h/J \lesssim 2(D/J - 1)$ and the uniaxial single-ion anisotropy approaches the value $D/J = 2$ from below. Under this condition, the transition lines gradually develop a pronounced S-shaped profile with initial bending toward lower fields and subsequent turning toward higher fields. As a result, two successive (reentrant) thermal phase transitions emerge in a relatively narrow interval of magnetic fields  [Figs.~\ref{fig:reentrant}(b)-(d)]. In this regime, the FRI phase is stable only within an intermediate temperature range and is enclosed by the DP phase at both lower and higher temperatures due to a sequence of DP-FRI-DP reentrant thermal phase transitions. Finally, the transition lines shown for $D/J=2$ [Fig.~\ref{fig:reentrant}(e)] and $D/J=2.5$ [Fig.~\ref{fig:reentrant}(f)] recover a conventional monotonic dependence, but now they bend toward higher magnetic field. 

\section{Monte Carlo simulations}
\label{mc}

To provide an independent verification of the exact analytical results reported above for the thermal phase transitions and simultaneously obtain a direct evidence for the spin arrangement realized in the individual phases, we performed classical Monte Carlo simulations of the effective spin-1/2 Ising model on a square lattice. According to the exact mapping relations (\ref{eq:m_sigma_definition}) and (\ref{eq:m_S_final}), we extracted from these Monte Carlo simulations the single-site magnetization of the nodal and the decorating spins of the corresponding mixed spin-$1/2$ and spin-$1$ Ising model on the Lieb lattice. To be more specific, the classical Monte Carlo simulations of the effective spin-1/2 Ising model were carried out using the single spin-flip Metropolis algorithm implemented within the Algorithms and Libraries for Physics Simulations (ALPS) project \cite{ALPS} for a square lattice with the total number of spins $14 400$, which corresponds to the mixed-spin Ising model on the Lieb lattice with a total number of $43 200$ spins.

\begin{figure}[!ht]
\centering
\includegraphics[width=0.49\linewidth]{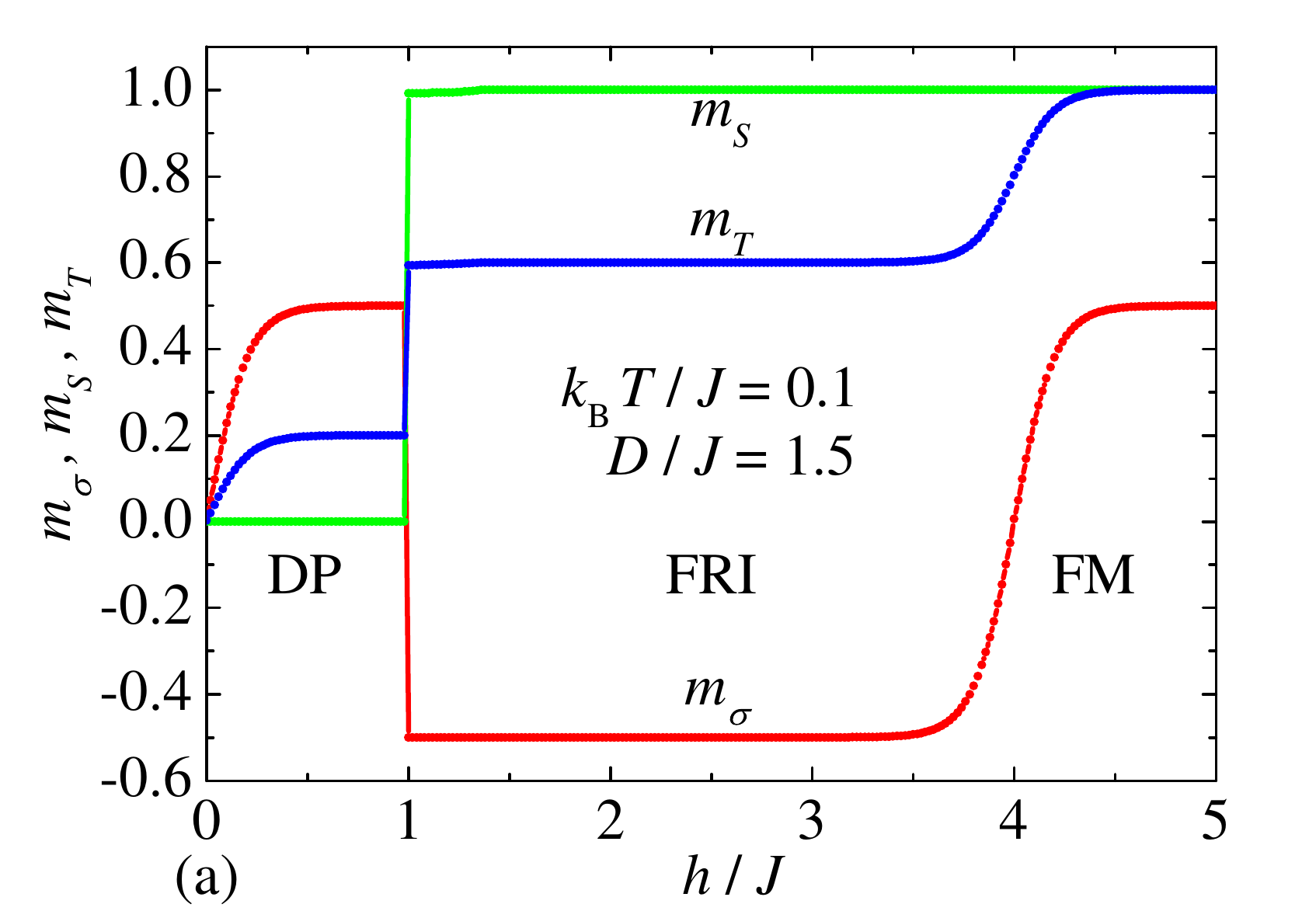}
\includegraphics[width=0.49\linewidth]{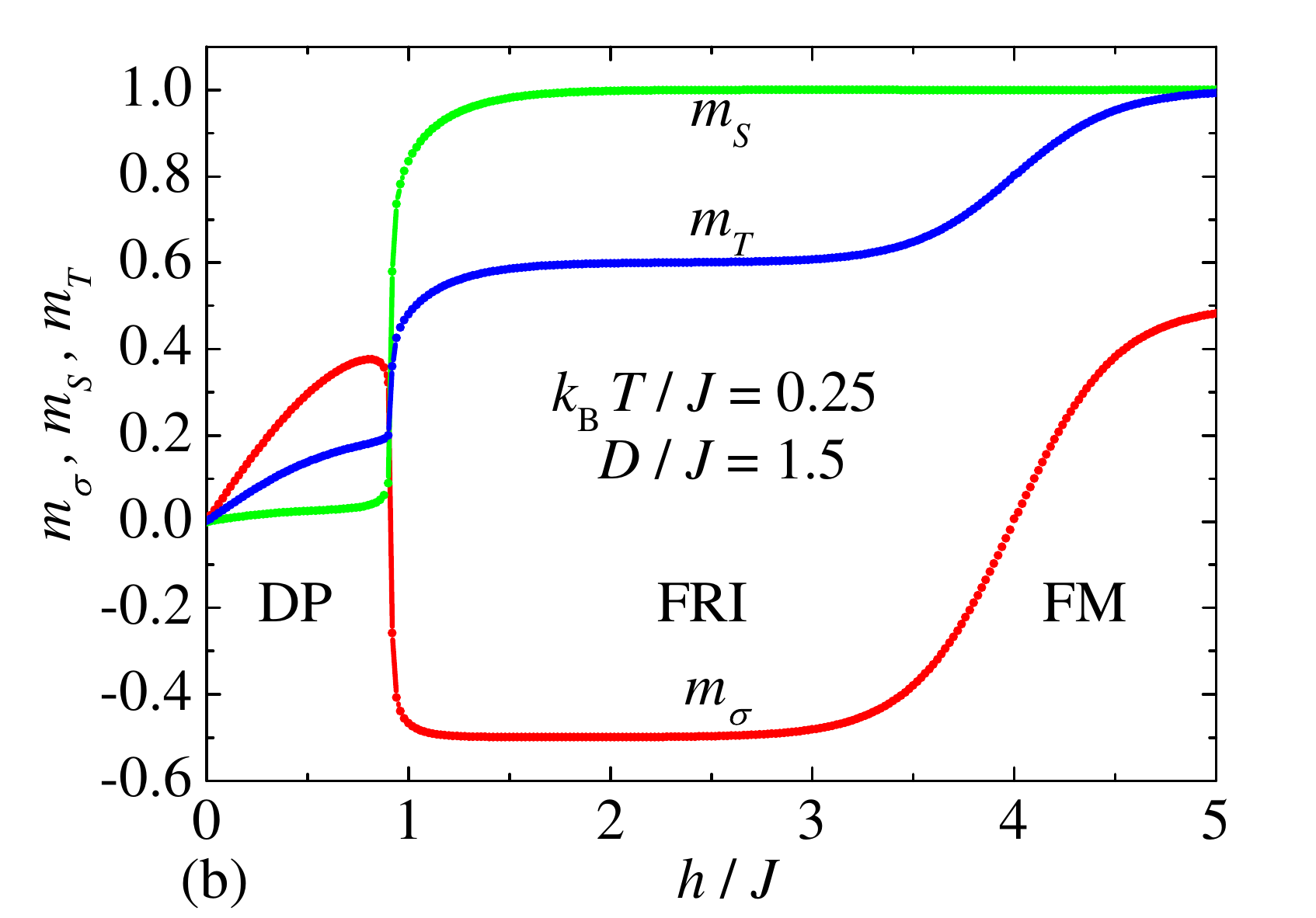}
\includegraphics[width=0.49\linewidth]{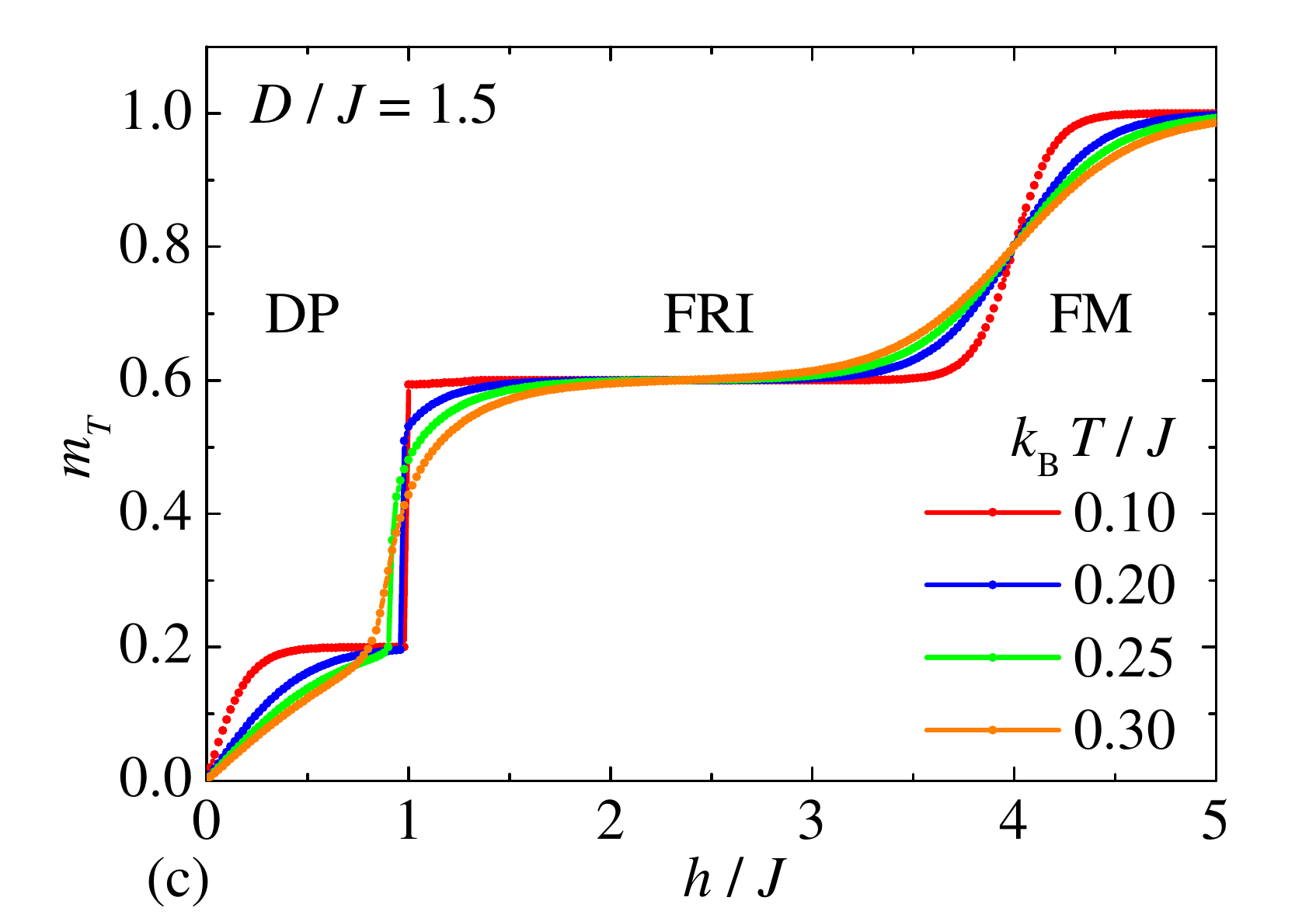}
\includegraphics[width=0.49\linewidth]{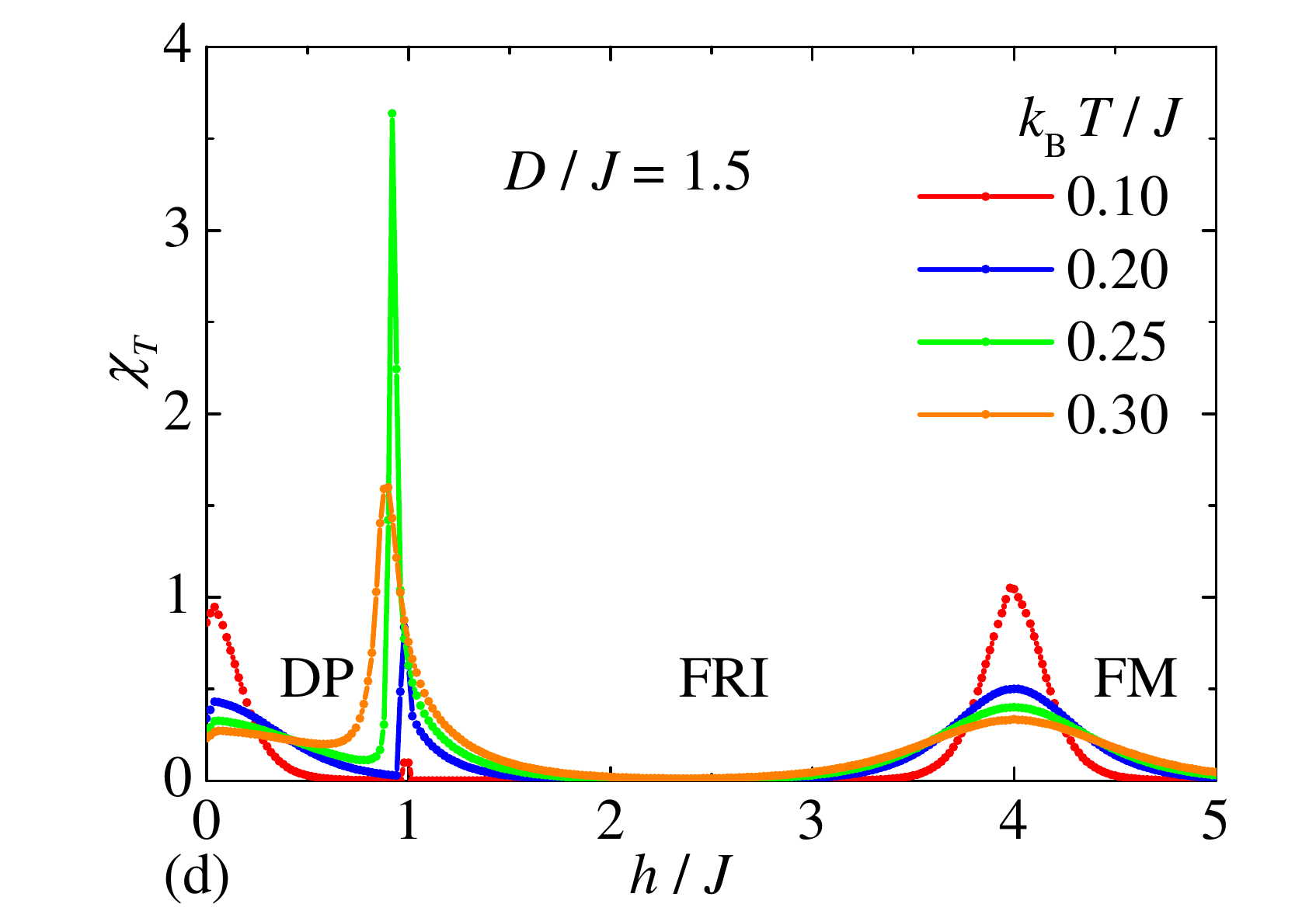}
\caption{Isothermal field dependencies of the magnetization and magnetic susceptibility of the mixed spin-$1/2$ and spin-$1$ Ising model on the Lieb lattice as obtained from classical Monte Carlo simulations for the fixed value of the uniaxial single-ion anisotropy $D/J=1.5$. The panels (a) and (b) show the single-site magnetization of the nodal ($m_\sigma$) and decorating ($m_S$) spins together with the total magnetization ($m_T$) normalized with respect to its saturation value at two selected temperature $k_{\rm B}T/J=0.1$ and $0.25$. The panels (c) and (d) display the total magnetization and magnetic susceptibility for four selected temperatures.}
\label{fig:1c5}
\end{figure}

First, we consider the mixed-spin Ising model on the Lieb lattice with the fixed value of the uniaxial single-ion anisotropy $D/J=1.5$, which serves as representative case with successive realization of the DP, FRI, and FM phases with increasing magnetic field. The magnetic-field dependencies of the single-site magnetization of the nodal and decorating spins $m_\sigma$ and $m_S$ are plotted in Fig.~\ref{fig:1c5}(a)-(b) together with the total magnetization $m_T$ normalized with respect to its saturation value. The isothermal magnetization curve displayed in Fig.~\ref{fig:1c5}(a) for the lowest temperature $k_{\rm B}T/J=0.1$ is characterized by an initial continuous increase of the local magnetization of the nodal spins from $m_\sigma=0$ toward its saturation value $m_\sigma=0.5$, whereas the local magnetization of the decorating spins stays nearly zero $m_S \simeq 0$. This behavior is fully consistent with the character of the DP phase, whose total magnetization is approximately $m_T\simeq0.2$. At the transition field $h/J \approx 1$, the system undergoes a discontinuous field-induced phase transition from the DP phase to the FRI phase. At this field-driven phase transition, the local magnetization of the decorating spins suddenly jumps from zero $m_S \simeq 0$ to the nearly saturated value $m_S \simeq 1$, while the local magnetization of the nodal spins contrarily undergoes an abrupt downturn from $m_\sigma \simeq 0.5$ to $m_\sigma \simeq -0.5$. Consequently, the total magnetization exhibits a discontinuous magnetization jump from the intermediate 1/5-plateau ($m_T \simeq 0.2$) characteristic of the DP phase to the intermediate 3/5-plateau ($m_T \simeq 0.6$) characteristic of the FRI phase. A further increase of the magnetic field leaves both local magnetization almost unchanged over a broad field interval corresponding to the FRI phase. Only when the magnetic field becomes sufficiently strong $h/J \approx 4$, the nodal spins start to align with the magnetic-field direction, which results in a smooth increase of the respective local magnetization from $m_\sigma=-0.5$ to $m_\sigma=0.5$. The accompanying increase of the total magnetization from $m_T \simeq 0.6$ to its saturation value $m_T=1$ signals a gradual thermally-assisted crossover from the FRI phase to the fully polarized FM phase. In contrast to the low-field DP-FRI phase transition, the high-field
FRI-FM crossover is not associated with a genuine phase transition.

The isothermal magnetization curve shown in Fig.~\ref{fig:1c5}(b) for the higher temperature $k_{\rm B}T/J=0.25$ displays a qualitatively different thermal phase transition, because the selected temperature is close to the terminal critical point of the DP-FRI phase-transition line. As a result, the discontinuous magnetization jump diminishes and eventually evolves into a critical point associated with a continuous field-driven phase transition. The overall temperature evolution of the magnetization curves is illustrated in Fig.~\ref{fig:1c5}(c), which shows the total magnetization for four representative temperatures. The magnetization curves at the two lowest temperatures $k_{\rm B}T/J=0.1$ and $0.2$ indicate a gradual temperature-induced melting of the discontinuous DP-FRI phase transition evidenced in a progressive reduction of the respective magnetization discontinuity. Upon approaching the critical temperature $k_{\rm B}T/J\approx 0.25$, the magnetization jump diminishes and changes into a singular critical point associated with a continuous field-driven phase transition. The isothermal magnetization curve at even higher temperature $k_{\rm B}T/J= 0.3$ is free of any phase transition and indicates a simpler thermally-assisted crossover between the DP and FRI phases. A similar crossover phenomenon is also uncovered in the magnetization curves at higher magnetic field $h/J = 4$, but this high-field crossover between the FRI and FM phases persists for any nonzero temperature unlike the low-field DP-FRI transition. 

A distinction between a crossover phenomenon, discontinuous and continuous phase transitions becomes particularly evident from the field variations of the magnetic susceptibility shown in Fig.~\ref{fig:1c5}(d). At the low temperatures $k_{\rm B}T/J=0.1$ and $0.2$, the discontinuous field-driven transition between the DP and FRI phases located at the transition field $h/J \approx 1$ manifests itself through a cusp-like anomaly in the magnetic susceptibility, whose amplitude is progressively enhanced with increasing temperature. Moreover, this susceptibility peak is expected to diverge when approaching the critical temperature and the pronounced peak observed in the magnetic susceptibility at $k_{\rm B}T/J=0.25$ represents a clear signature of the close proximity of the critical point. The susceptibility finally displays a round maximum above the critical temperature (e.g., $k_{\rm B}T/J=0.3$), which gradually decreases in magnitude with further increasing temperature. By contrast, the smooth crossover behavior between the FRI and FM phases is always accompanied by a round maximum of the magnetic susceptibility, which can be observed at the higher magnetic field $h/J \approx 4$ for arbitrary nonzero temperature. In general, this crossover-induced round maximum becomes progressively broader and lower as the temperature increasing.

\begin{figure}[!ht]
\centering
\includegraphics[width=0.49\linewidth]{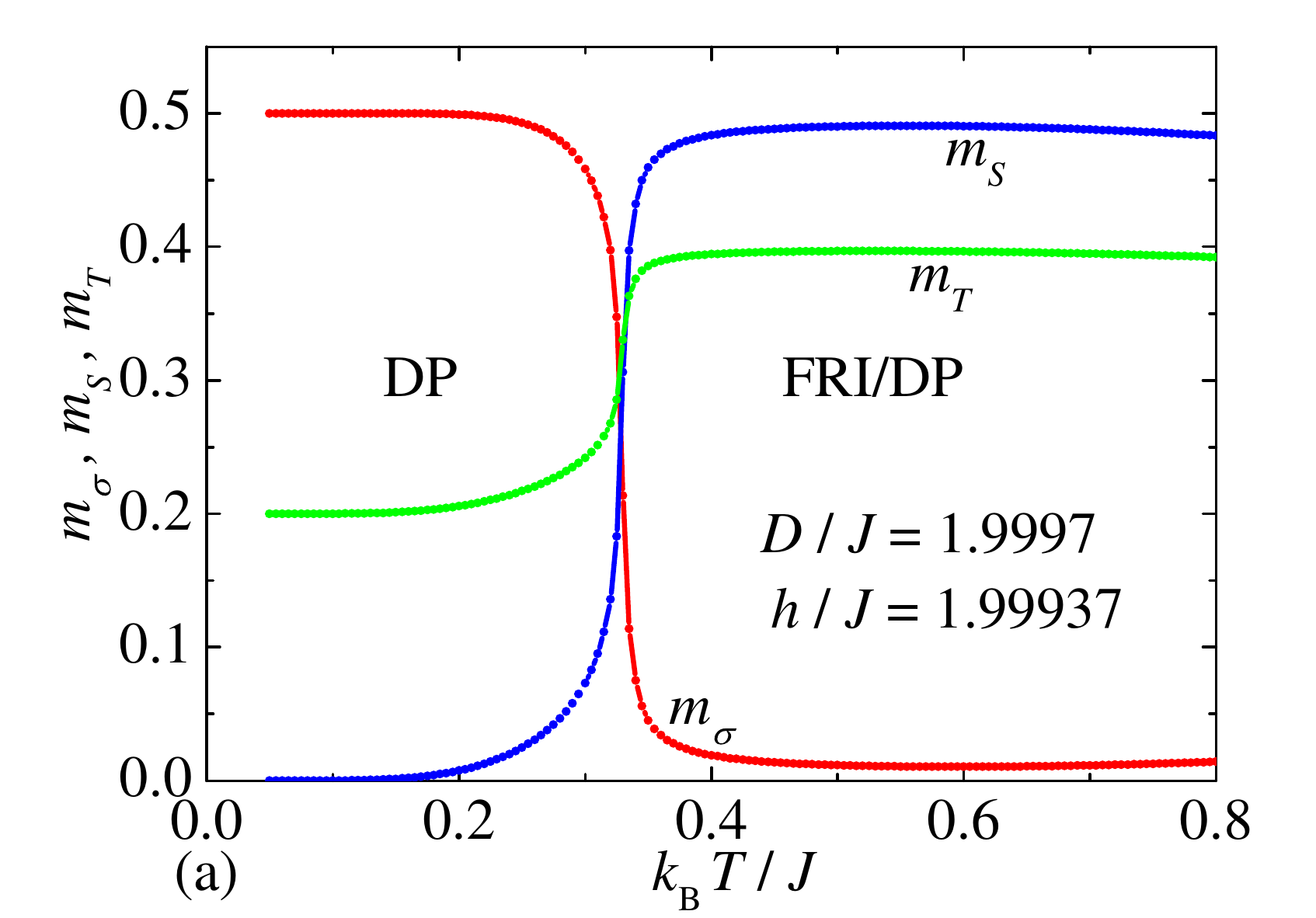}
\includegraphics[width=0.49\linewidth]{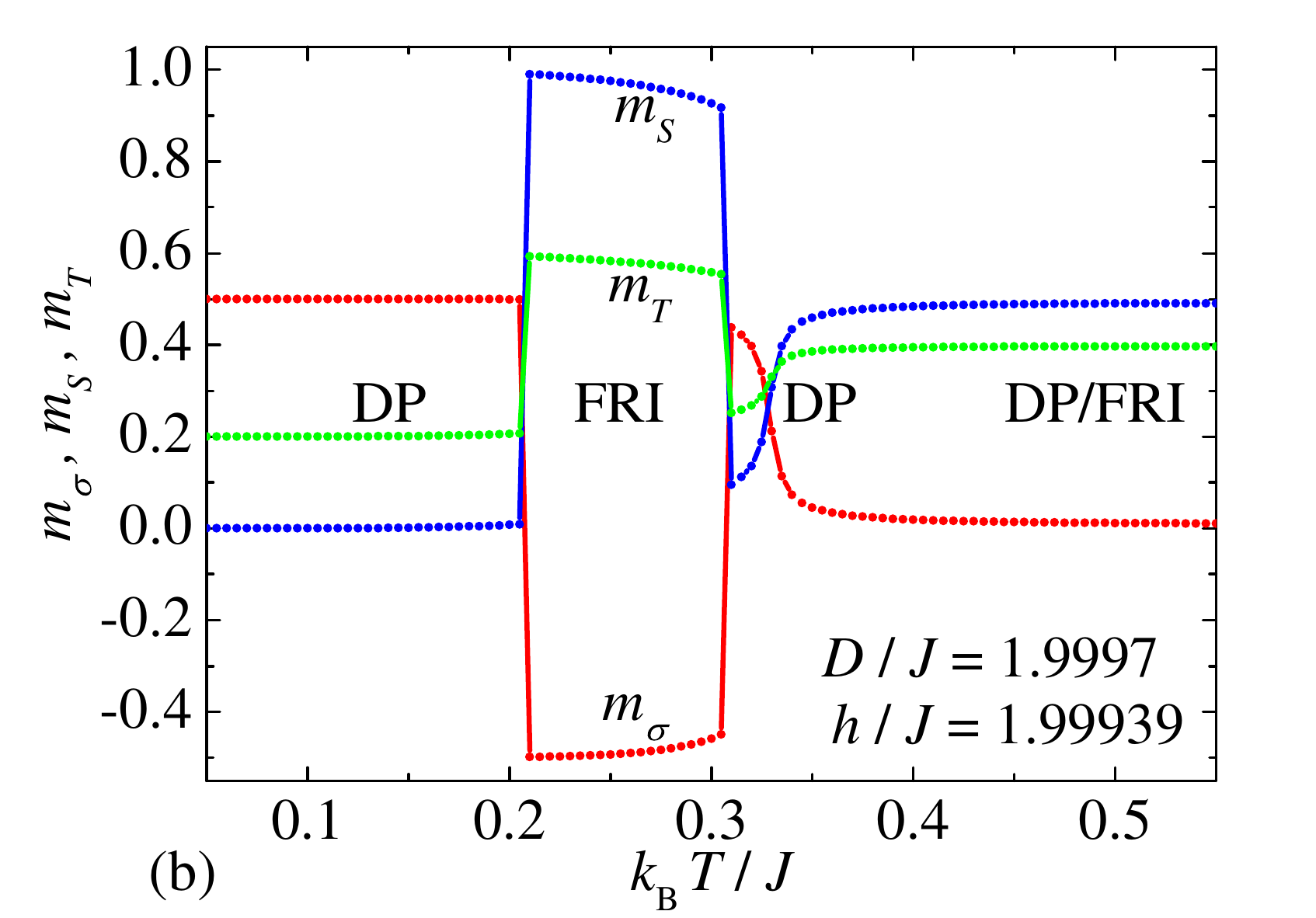}
\includegraphics[width=0.49\linewidth]{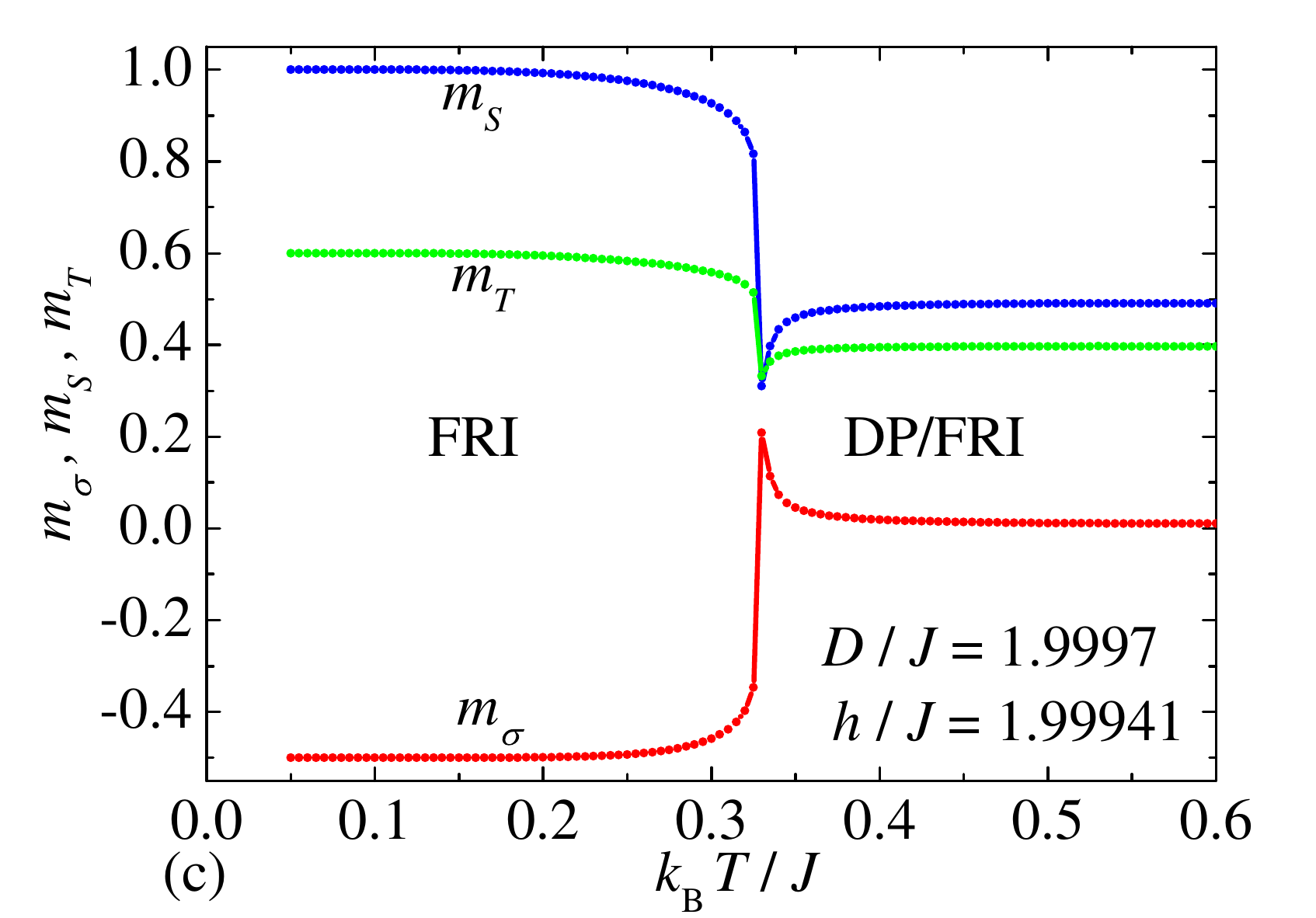}
\includegraphics[width=0.49\linewidth]{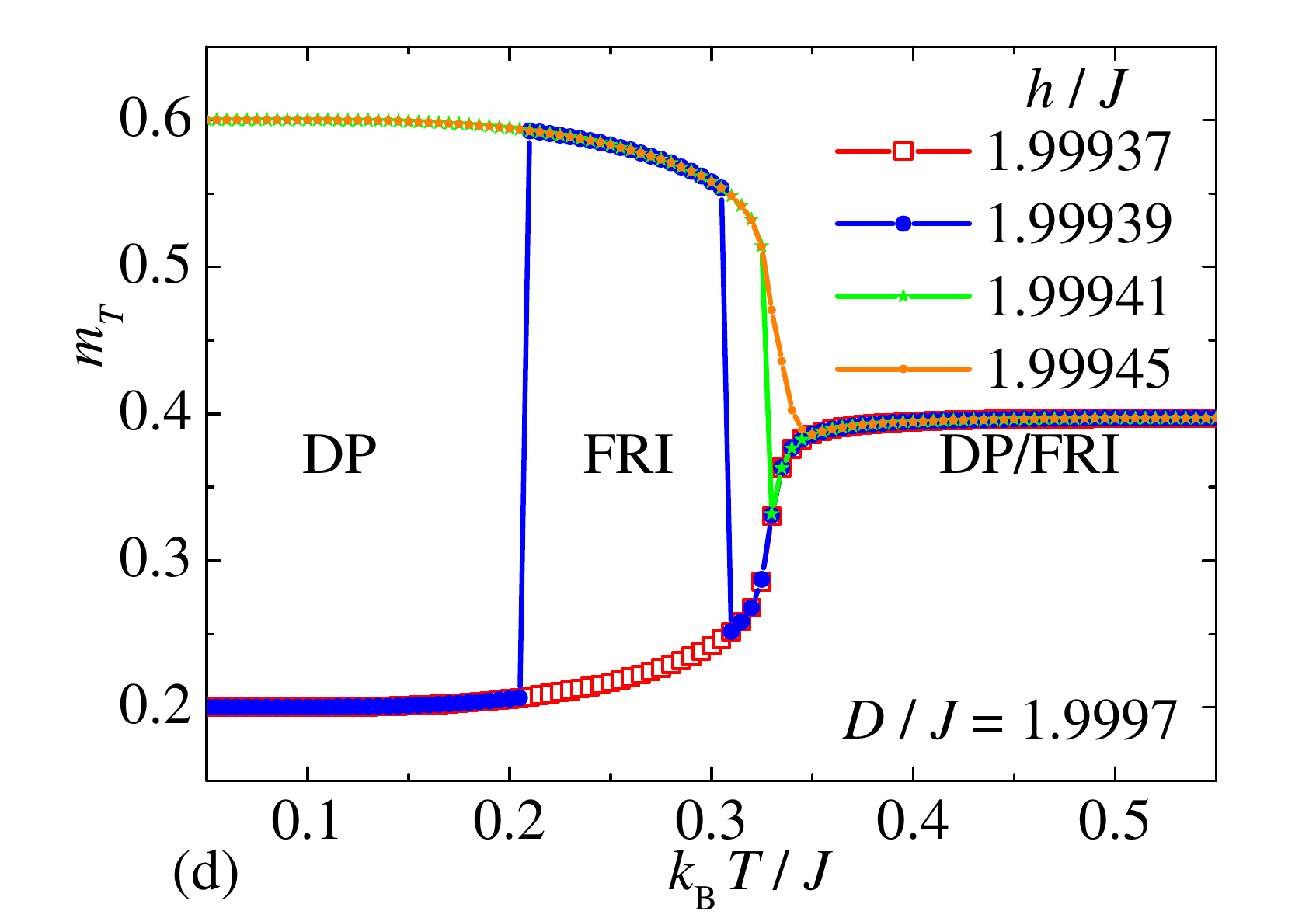}
\caption{Temperature dependencies of the local and total magnetization of the mixed spin-$1/2$ and spin-$1$ Ising model on the Lieb lattice as obtained from the classical Monte Carlo simulations for the fixed value of the uniaxial single-ion anisotropy $D/J=1.9997$ and three selected values of the magnetic field: (a) $h/J=1.99937$, (b) $h/J=1.99939$,(c) $h/J=1.99941$. The panel (d) presents temperature dependence of the total magnetization $m_T$ for four representative values of the magnetic field.}
\label{fig:tep}
\end{figure}

Next, we use classical Monte Carlo simulations to corroborate reentrant thermal phase transitions of the mixed spin-$1/2$ and spin-$1$ Ising model on the Lieb lattice, which were predicted in the previous section on the grounds of exact analytical results in a very narrow parameter regime for the uniaxial single-ion anisotropy $D/J \lesssim 2$. To this end, we fix a relative strength of the uniaxial single-ion anisotropy to the specific value $D/J=1.9997$, which should according to the finite-temperature phase diagram shown in Fig. \ref{fig:reentrant}(c) support existence of reentrant thermal phase transitions for a suitably chosen magnetic-field strength. The local magnetization of the nodal and decorating spins $m_\sigma$ and $m_S$ are plotted in Fig.~\ref{fig:tep} together with the total magnetization $m_T$ for several magnetic fields. For the magnetic field $h/J=1.99937$ [Fig.~\ref{fig:tep}(a)], both low-temperature values of the local magnetization $m_\sigma \simeq 0.5$ and $m_S \simeq 0$ are consistent with the spin arrangement of the DP phase, which undergo a marked but still smooth crossover change around the temperature $k_{\rm B}T/J \approx 0.33$ toward the specific values $m_\sigma \simeq 0$ and $m_S \simeq 0.5$ typical for a phase coexistence of the DP and FRI phases. As a consequence, the total magnetization evolves smoothly without any singular behavior from the value characteristic of the DP phase to a rather specific average value of the FRI and DP phases.

Even more striking temperature variations of the magnetization are observed in Fig.~\ref{fig:tep}(b) for the magnetic field $h/J=1.99939$, where clear signatures of reentrant thermal phase transitions between the DP and FRI phases emerge. At low enough temperatures, the system resides in the DP phase characterized by the local magnetization of the nodal and decorating spins $m_\sigma \simeq 0.5$ and $m_S \simeq 0$ resulting in the total magnetization $m_T \simeq 0.2$. These values of the local and total magnetization remain essentially unchanged up to the temperature $k_{\rm B}T/J \approx 0.21$, at which a discontinuous thermally-induced phase transition takes place. Immediately above this transition temperature, the local magnetization of the decorating spins abruptly jumps to a nearly saturated value $m_S \simeq 1$, while the local magnetization of the nodal spins undergoes a sudden change to the specific value $m_\sigma \simeq -0.5$ consistently with the spin arrangement of the FRI phase characterized by the total magnetization $m_T \simeq 0.6$. A further increase of temperature leads to a second discontinuous thermally-induced phase transition at $k_{\rm B}T/J \approx 0.31$, above which the system reenters the DP phase as evidenced by the typical values of the local and total magnetization. Finally, the local and total magnetization tend at even higher temperatures $k_{\rm B}T/J \gtrsim 0.35$ to the specific average values, which are characteristic for a phase coexistence of the FRI and DP phases. Altogether, the Monte Carlo results provide a direct numerical evidence for reentrant thermal phase transitions DP$\rightarrow$FRI$\rightarrow$DP arising from the intermediate-temperature stabilization of the FRI phase already predicted by the exact analytical solution.

The temperature dependencies of the local and total magnetization shown in Fig.~\ref{fig:tep}(c) for the specific value of the magnetic field $h/J=1.99941$ corresponds to the close vicinity of the terminal critical point of the line of discontinuous phase transition between the FRI and DP phases. Under this specific condition, the low-temperature values of the local magnetization of the nodal and decorating spins $m_\sigma \simeq -0.5$ and $m_S \simeq 1$ are characteristic of the FRI phase and magnetization discontinuities vanish. At the critical temperature $k_{\rm B}T/J \approx 0.33$, both local magnetization display a steep but still continuous power-law dependence, which is subsequently followed by a tail-like behavior as the local and total magnetization gradually approach average values typical for a phase coexistence of the FRI and DP phases. A comprehensive overview of these temperature-induced changes is provided by Fig.~\ref{fig:tep}(d), which compares the temperature dependence of the total magnetization for several representative magnetic fields. For the lowest field $h/J=1.99937$, the total magnetization changes smoothly from the low-temperature DP value $m_T\simeq0.2$ to the high-temperature value $m_T\simeq0.4$ confirming the absence of a genuine phase transition. A slight increase of the magnetic field to $h/J=1.99939$ gives rise to a well-developed intermediate FRI phase with $m_T\simeq0.6$, which is separated from the low- and high-temperature DP phase by two discontinuous phase transitions. For $h/J=1.99941$, the temperature interval over which the FRI phase remains stable becomes noticeably narrower reflecting the close proximity of the terminal critical point. Finally, the discontinuities disappear completely for $h/J=1.99945$ when the magnetization varies continuously with temperature.  These Monte Carlo results thus provide direct numerical evidence for the temperature-induced reentrant behavior predicted by the exact analytical solution and demonstrate its gradual disappearance upon approaching the terminal critical point.

\section{Similarity with the Ising-Heisenberg diamond-decorated square lattice}
\label{sim}

A remarkable similarity can be identified between thermal phase transitions of the present mixed spin-1/2 and spin-1 Ising model on the Lieb lattice and those of the spin-$1/2$ Ising-Heisenberg model on a diamond-decorated square lattice studied in Ref. \cite{Strecka2023}. From the microscopic point of view, the spin-$1/2$ Ising-Heisenberg model on a diamond-decorated square lattice differs from the mixed spin-1/2 and spin-1 Ising model on the Lieb lattice in that it involves on each bond of a square lattice one couple of the quantum Heisenberg spins instead of a classical spin-1 Ising variable. A sufficiently strong antiferromagnetic coupling $J_2$ within the Heisenberg spin pairs favors formation of the dimer-singlet state, which effectively decouples the nodal Ising spins quite similarly as the nonmagnetic state $S_i=0$ of the decorating spins does in the mixed-spin Ising model on the Lieb lattice. In this sense, the effect of the easy-plane single-ion anisotropy closely resembles that of antiferromagnetic coupling within the Heisenberg dimers. Moreover, the two magnetic states $S_i = \pm 1$ of the decorating spins closely resemble two polarized components dimer-triplet states of the Heisenberg spin pairs. The only fundamental difference between the two models thus lies in a nonmagnetic component of the triplet-dimer state of the Heisenberg dimers in the spin-$1/2$ Ising-Heisenberg model on a diamond-decorated square lattice, which has no counterpart in the mixed spin-1/2 and spin-1 Ising model on the Lieb lattice.  

Despite their substantially different microscopic structures, both models can be rigorously mapped by means of the generalized decoration-iteration transformation onto an effective spin-$1/2$ Ising model on a square lattice characterized by the effective interaction $J_{\rm eff}$ and the effective field $h_{\rm eff}$. Upon identifying the connection $D \leftrightarrow J_2$ between the uniaxial single-ion anisotropy $D$ and the intra-dimer interaction $J_2$ within the Heisenberg spin pairs, the analogy becomes not only qualitative but also quantitative as evidenced by the comparison of the effective interaction $J_{\rm eff}$ and the effective field $h_{\rm eff}$ given by Eqs. (\ref{eq:Jeff})-(\ref{eq:v123}) of the present paper and Eqs. (8)-(10) of Ref. \cite{Strecka2023}. These effective mapping parameters differ solely through the contribution of a nonmagnetic component of the triplet-dimer state of the Heisenberg dimers. Taking into consideration a mapping correspondence to the effective Ising model with almost the same effective interaction and effective field, both models consequently exhibit very similar ground-state and finite-temperature phase diagrams including a narrow parameter region hosting thermally-induced reentrant phase transitions. This similarity can be readily appreciated by comparing Figs. \ref{gspd1} and \ref{fig:reentrant} of the present work with Figs. 4 and 8 of Ref. \cite{Strecka2023}. 

\section{Conclusion}
\label{conc}

In this paper, we have exactly investigated discontinuous and continuous thermal phase transitions of the mixed spin-$1/2$ and spin-$1$ Ising model on the Lieb lattice accounting for the uniaxial single-ion anisotropy and the external magnetic field. By applying the generalized decoration-iteration transformation, we established an exact mapping correspondence with an effective spin-$1/2$ Ising model on the square lattice. This mapping enabled us to determine rigorously the ground-state and finite-temperature phase diagrams including the lines of discontinuous thermal phase transitions terminating at Ising-type critical points associated with continuous thermal phase transitions.

The ground-state analysis reveals the existence of three different FRI, DP and FM phases. Although all ground-state phase boundaries correspond to discontinuous field-driven phase transitions, discontinuous thermal phase transitions are confined to a dome-shaped wall evolving exclusively from the ground-state phase boundary between the FRI and DP phases. This surface of the discontinuous thermal phase transition is terminated by the line of continuous thermal phase transitions belonging to the two-dimensional Ising universality class. 

A particularly intriguing feature of of the mixed spin-$1/2$ and spin-$1$ Ising model on the Lieb lattice is the existence of a narrow parameter region $D/J \lesssim 2$ hosting reentrant thermal phase transitions. The mutual competition between the nearly balanced uniaxial single-ion anisotropy and the magnetic field gives rise to two successive discontinuous thermal phase transition with the sequence DP$\rightarrow$FRI$\rightarrow$DP achieved upon increasing the temperature. The exact analytical predictions for the thermal phase transitions including the reentrant ones were further corroborated by  classical Monte Carlo simulations, which provided precise numerical results fully consistent with these analytical predictions.

A comparison of the thermal phase transitions observed in the present mixed spin-$1/2$ and spin-$1$ Ising model on the Lieb lattice with those previously reported for the spin-1/2 Ising-Heisenberg model on the diamond-decorated square lattice \cite{Strecka2023} reveals a remarkable similarity. This similarity is not accidental, because both microscopically distinct spin systems can be mapped onto the same effective spin-1/2 Ising model on the square lattice with nearly identical effective interaction and effective field. In this respect, the present mixed spin-1/2 and spin-1 Ising model on the Lieb lattice provides valuable insight into the origin of discontinuous thermal phase transitions, Ising critical points, and reentrant phase transitions of more complex models such as the hybrid classical-quantum spin-$1/2$ Ising-Heisenberg model  \cite{Strecka2023} or even fully quantum spin-1/2 Heisenberg model \cite{Caci2023} on the diamond-decorated square lattice, for which obtaining exact analytical results remains elusive.

\section*{Acknowledgement}
The authors acknowledge funding by the grant of Slovak Research and Development Agency under the contract No. APVV-24-0091, by the grant of The Ministry of Education, Research, Development and Youth of the Slovak Republic under the contract No. VEGA 1/0298/25, and the EU NextGenerationEU through the Recovery and Resilience Plan for Slovakia under the project No. 09I03-03-V04-00403.

\end{document}